\documentclass[twocolumn,trackchanges]{aastex62}

\usepackage{mathtools}
\usepackage{CJKutf8}

\newcommand{\ra}[4]{$#1\overset{\mathrm h}{\phd}#2\overset{\mathrm m}{\phd}#3\overset{\mathrm s}{.}#4$}
\newcommand{\dec}[4]{$#1\overset{\circ}{\phd}#2\overset{\prime}{\phd}#3\overset{\prime\prime}{.}#4$}

\newcommand{\jp}[1]{~(\begin{CJK*}{UTF8}{min}#1\end{CJK*}\!\!)}
\newcommand{\msun}{M\textsubscript{\(\odot\)}}
\newcommand{\rsun}{R\textsubscript{\(\odot\)}}

\graphicspath{{./}{figures/}}

\received{Soon}
\revised{After that}
\accepted{After that}
\submitjournal{ApJ}

\shorttitle{The Transitional 02es-like SN 2019yvq}
\shortauthors{Burke et al.}

\begin{document}

\title{A Bright Ultraviolet Excess in the Transitional 02es-like Type Ia Supernova 2019yvq}

\correspondingauthor{J. Burke}
\email{jburke@lco.global}

\author[0000-0003-0035-6659]{J. Burke}
\affiliation{Department of Physics, University of California, Santa Barbara, CA 93106-9530, USA}
\affiliation{Las Cumbres Observatory, 6740 Cortona Dr, Suite 102, Goleta, CA 93117-5575, USA}

\author[0000-0003-4253-656X]{D. A. Howell}
\affiliation{Department of Physics, University of California, Santa Barbara, CA 93106-9530, USA}
\affiliation{Las Cumbres Observatory, 6740 Cortona Dr, Suite 102, Goleta, CA 93117-5575, USA}

\author[0000-0002-4781-7291]{S. K. Sarbadhicary}
\affil{Center for Data Intensive and Time Domain Astronomy, Department of Physics and Astronomy, Michigan State University, East Lansing, MI 48824}

\author[0000-0003-4102-380X]{D. J. Sand}
\affil{Steward Observatory, University of Arizona, 933 North Cherry Avenue, Tucson, AZ 85721-0065, USA}

\author[0000-0002-1546-9763]{R. C. Amaro}
\affil{Steward Observatory, University of Arizona, 933 North Cherry Avenue, Tucson, AZ 85721-0065, USA}

\author[0000-0002-1125-9187]{D. Hiramatsu}
\affiliation{Department of Physics, University of California, Santa Barbara, CA 93106-9530, USA}
\affiliation{Las Cumbres Observatory, 6740 Cortona Dr, Suite 102, Goleta, CA 93117-5575, USA}

\author[0000-0001-5807-7893]{C. McCully}
\affiliation{Department of Physics, University of California, Santa Barbara, CA 93106-9530, USA}
\affiliation{Las Cumbres Observatory, 6740 Cortona Dr, Suite 102, Goleta, CA 93117-5575, USA}

\author[0000-0002-7472-1279]{C. Pellegrino}
\affiliation{Department of Physics, University of California, Santa Barbara, CA 93106-9530, USA}
\affiliation{Las Cumbres Observatory, 6740 Cortona Dr, Suite 102, Goleta, CA 93117-5575, USA}

\author[0000-0003-0123-0062]{J. E. Andrews}
\affil{Steward Observatory, University of Arizona, 933 North Cherry Avenue, Tucson, AZ 85721-0065, USA}

\author[0000-0001-6272-5507]{P. J. Brown}
\affil{Department of Physics and Astronomy, Texas A\&M University, 4242 TAMU, College Station, TX 77843, USA}
\affil{George P. and Cynthia Woods Mitchell Institute for Fundamental Physics \& Astronomy}
 
\author{Koichi~Itagaki \jp{板垣公一}}
\affil{Itagaki Astronomical Observatory, Yamagata 990-2492, Japan}

\author[0000-0002-9301-5302]{M. Shahbandeh}
\affiliation{Department of Physics, Florida State University, Tallahassee, FL 32306, USA}





\author[0000-0002-4294-444X]{K.~A. Bostroem}
\affil{Department of Physics and Astronomy, University of California, 1 Shields Avenue, Davis, CA 95616-5270, USA}

\author[0000-0002-8400-3705]{L. Chomiuk}
\affil{Center for Data Intensive and Time Domain Astronomy, Department of Physics and Astronomy, Michigan State University, East Lansing, MI 48824}

\author[0000-0003-1039-2928]{E.~Y.  Hsiao}
\affiliation{Department of Physics, Florida State University, Tallahassee, FL 32306, USA}

\author[0000-0001-5510-2424]{Nathan Smith}
\affil{Steward Observatory, University of Arizona, 933 North Cherry Avenue, Tucson, AZ 85721-0065, USA}

\author[0000-0001-8818-0795]{S. Valenti}
\affiliation{Department of Physics and Astronomy, University of California, 1 Shields Avenue, Davis, CA 95616-5270, USA}








\begin{abstract}

\noindent
We present photometric and spectroscopic observations of the nearby Type Ia SN 2019yvq,
from its discovery $\sim$1 day after explosion to $\sim$100 days after its peak brightness.
This SN exhibits several unusual features,
most notably an extremely bright UV excess seen within $\sim$5 days of its explosion.
As seen in \textit{Swift} UV data, this early excess outshines its ``peak" brightness,
making this object more extreme than other SNe with early UV/blue excesses (e.g. iPTF14atg and SN 2017cbv).
In addition, it was underluminous ($M_B=-18.4$),
relatively quickly declining ($\Delta m_{15}(B)=1.35$),
and shows red colors past its early blue bump.
Unusual (although not unprecedented) spectral features include extremely broad-lined and high-velocity Si absorption.
Despite obvious differences in peak spectra,
we classify SN 2019yvq as a transitional member of the 02es-like subclass due to its similarities in several respects 
(e.g. color, peak luminosity, peak Ti, nebular [Ca II]).
We model this dataset with a variety of published models,
including SN ejecta--companion shock interaction and sub-Chandrasekhar mass WD double detonation models.
Radio constraints from the VLA place an upper limit of $(4.5 \text{---} 20) \times 10^{-8}$ M$_{\odot}$/yr on the mass-loss rate from a symbiotic progenitor,
which does not exclude a red giant or main sequence companion.
Ultimately we find that no one model can accurately replicate all aspects of the dataset,
and further we find that the ubiquity of early excesses in 02es-like SNe Ia requires a progenitor system that is capable of producing isotropic UV flux,
ruling out some models for this class of objects.

\end{abstract}


\section{Introduction} \label{sec:intro}

Despite the fact that Type Ia supernovae (SNe) were used as standardizable candles to discover the accelerating expansion of the universe and constrain its energy content \citep{Riess_1998, Perlmutter_1999}, open questions remain about their progenitor systems. 
The SNe themselves are understood to be the thermonuclear explosions of carbon/oxygen white dwarfs (WDs) \citep{hoyle}, but beyond that there are large uncertainties about both the progenitor system(s) and explosion mechanism(s).

Many possible progenitor systems have been theorized. 
The two broad classes are the single-degenerate channel \citep{whelan}, 
where the WD accretes matter slowly from a nondegenerate companion, 
and the double-degenerate channel \citep{iben},
where the source of the extra matter needed to ignite the WD is a second WD. 
Within these two broad channels exist many specific and sometimes exotic scenarios,
e.g. dynamically driven double-degenerate double-detonation systems \citep{shen6d} or rotating super-Chandrasekhar mass WD progenitors \citep{yoon}. 
For reviews, see \citet{howell11}, \citet{wanghan}, and \citet{maoz}.

\citet{kasen10} predicted an observational signature that could distinguish between the single- and double-degenerate cases.
If the donor star were nondegenerate then the SN ejecta will run into it and get shock-heated.
The shock-heated ejecta would then emit an excess of UV/blue light which could be detected in the SN's early-time lightcurve.
The strength of this signature is dependent on the companion's size and separation, the velocity of the ejecta, and the viewing angle of the event.
\citet{kasen10} predicted that the viewing angle effect alone would make this early blue excess visible in only 10\% of SNe Ia which explode through this single-degenerate channel.

Following the publication of \citet{kasen10}, many rolling supernova searches were examined for evidence of the effect in the optical and UV \citep{Hayden_etal_2010_shock, Bianco_etal_2011, ganeshalingam_powerlawrise, tucker_2011}.
These found no evidence for the predicted shock with a red giant companion.
\citet{Brown_etal_2012_shock} also excluded red giant companions from a smaller sample of SNe Ia with constraining UV data.
The early optical observations of SN 2011fe were additionally able to place extremely tight constraints on optical and UV shock emission from the companion \citep{Nugent11, Brown_etal_2012_11fe}.

Early blue excesses have since been seen in a small number of SNe,
most notably SN 2012cg \citep{Marion16},
iPTF14atg \citep{Cao15},
iPTF16abc \citep{Miller18},
and SN 2017cbv \citep{Hosseinzadeh17}.
The proliferation of transient surveys has allowed for much more consistent and thorough followup of young SNe \citep[e.g.][]{yao_ztf1}.
This in turn has revealed a wide range of early behaviors
including varying early color evolution
\citep{bulla_ztf3,redvsblue,brown_red,brown_uvcolors}
and a range of (sometimes broken) power laws which describe their rising lightcurves
\citep{Olling15,Miller18,miller_ztf2,li_18oh,shappee_18oh,dimitriadis_18oh}.

A number of progenitor scenarios can reproduce some range of these observed properties,
including explosions which vary the degree of nickel mixing in the exploding WD \citep{Piro16} leading to a range of early colors,
and models of sub-Chandrasekhar mass WDs detonated by the ignition of a surface layer of He \citep{polin_subch} leading to a wide range of absolute magnitudes and colors.

In this paper we present early-time photometry and spectroscopy of the Type Ia SN 2019yvq,
a SN discovered in late 2019 which displays a rare, and unusually strong, blue bump at early times.
The object displays other unusual behavior, including extremely broad and high-velocity Si II at peak and strong nebular [Fe II] and [Ca II].
Its unique combination of characteristics make it an excellent stress-test for several models of SNe Ia.
Multiple papers have already been written about this object \citep{miller2020_19yvq,siebert_19yvq,tucker_19yvq},
which we reference throughout,
as this work agrees with prior findings in some respects and disagrees in others.

In Section \ref{sec:obs} we describe the object's discovery and the observational followup by 
Las Cumbres Observatory,
which obtained data presented here for the first time,
and the \textit{Swift} space telescope.
In Section \ref{sec:data_analysis} we discuss interesting features of the dataset,
and we compare specifically to 02es-like SNe Ia in Section \ref{sec:02es_comp}.
In Section \ref{sec:models} we compare our data to models from \citet{kasen10} and \citet{polin_subch} and discuss the difficulty of finding a single model that reproduces all features of our dataset.
In Section \ref{sec:radio} we discuss constraints on the progenitor system as indicated by radio observations from the Karl G. Jansky Very Large Array.
We discuss implications of the event and its properties in Section \ref{sec:discussion}.
We conclude in Section \ref{sec:conclude}.



\section{Discovery \& Observations} \label{sec:obs}

\subsection{Discovery}\label{sec:discovery}

SN~2019yvq was discovered by Koichi Itagaki \citep{SN19yvq_discovery} on 2019 December 28.74 UT using a Celestron 14 inch telescope at an unfiltered magnitude of 16.7.
A nondetection of the same field, using an identical setup, was found the night before (2019 December 27.72 UT), with a limiting unfiltered magnitude of $\sim$18.2.
This nondetection is approximately 0.3 days after the nondetection reported by ASAS-SN in \citet{tucker_19yvq},
and places an even more stringent limit on the rise-time and early lightcurve.
Following the initial discovery, both the ZTF \citep{ztf} and ATLAS \citep{tonry_atlas} surveys reported detections of SN~2019yvq.
An initial classification spectrum using HOWPol on the 1.5-m Kanata telescope on 2020 January 01.84 suggested that SN~2019yvq was a Type Ib/c supernova \citep{SN19yvq_class1}, although a subsequent spectrum (taken on 2020 January 4.07) with the SPRAT spectrograph on the Liverpool telescope clearly showed that SN2019yvq was a SN Ia before maximum light.
A spectrum from the SED Machine on the Palomar 60-in telescope taken on 2020 January 12.36
further confirmed that SN 2019yvq is a SN Ia.
We have downloaded these spectra from the Transient Name Server (TNS)\footnote{\url{https://wis-tns.weizmann.ac.il/}} and incorporated them into our analysis.

\begin{figure}
\begin{center}
\includegraphics[width=0.5\textwidth]{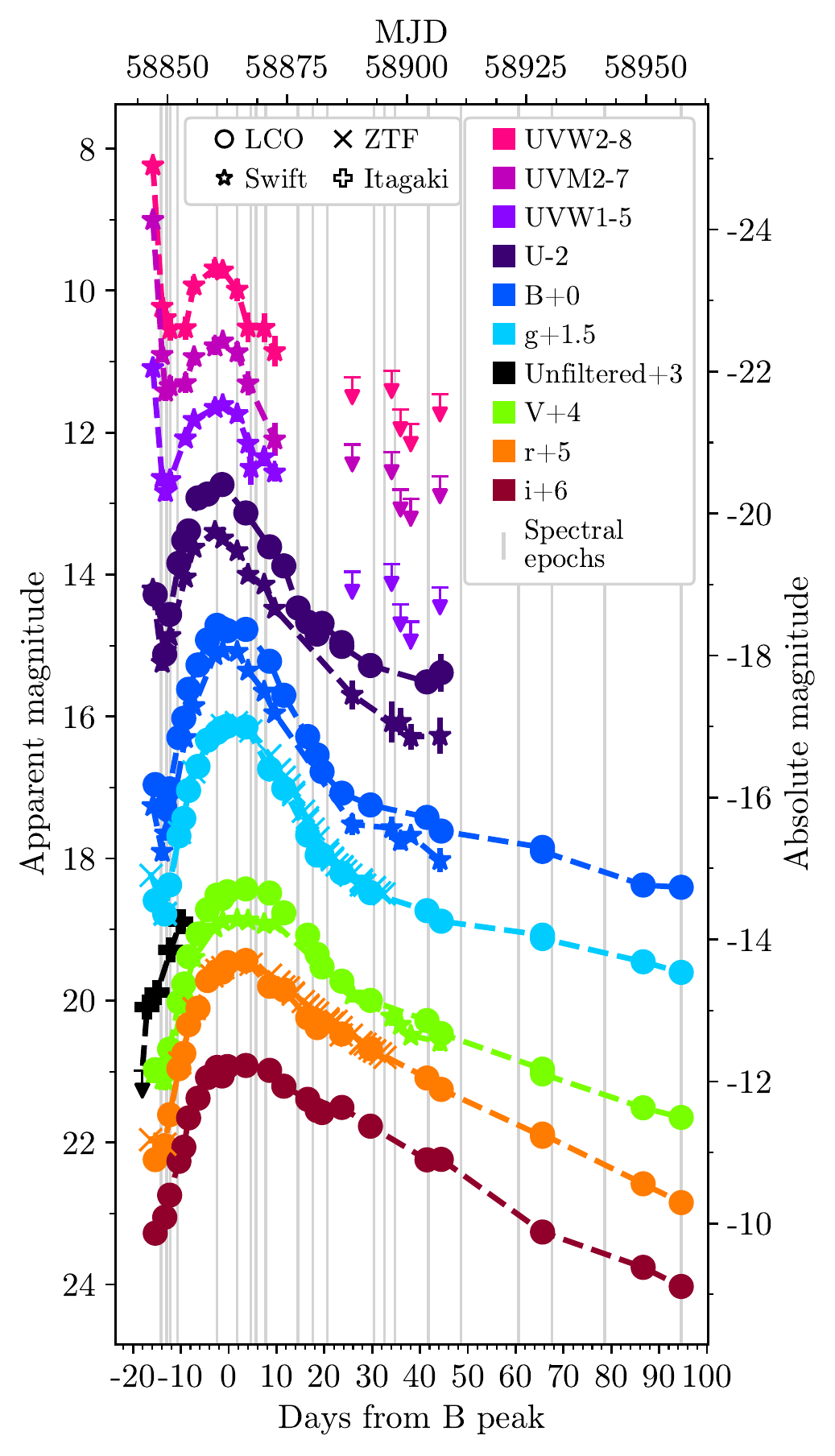}
\caption{ UV and optical extinction-corrected photometry of SN 2019yvq.
As discussed in Section \ref{ssec:lightcurve_analysis} we adopt $E(B-V)_{\textrm{host}}=0.052$ throughout our analysis,
in addition to $E(B-V)_{\textrm{Milky Way}}=0.017$.
The first epoch shows an extremely strong blue/UV excess. 
The lines connecting the points are simple linear interpolations to guide the eye, especially to the strength of the early UV excess, and do not represent models.
The epochs of the spectra shown in Figure \ref{fig:spec_timeseries} are included as vertical grey lines.
\label{fig:lc}}
\end{center}
\end{figure}

SN 2019yvq is located at right ascension \ra{12}{27}{21}{85} and declination \dec{+64}{47}{59}{8} (J2000), and lies 12.9 arcsec to the southeast of the host galaxy NGC 4441, which has a redshift of $z$=0.00908 \citep[][retrieved via NED\footnote{\url{http://ned.ipac.caltech.edu/}}]{NGC4441_redshift}.
NGC 4441 is an SAB0-type galaxy, and is clearly undergoing a merger event as can be seen in deep images from the DESI Legacy Imaging Survey\footnote{\url{http://legacysurvey.org/viewer}} \citep{DESI_imaging}.
A surface brightness fluctuation (SBF) distance to NGC 4441 suggests $D$$\approx$20 Mpc \citep{Tonry01},
although the disturbed nature of the host likely affects this measurement.
The Hubble-flow distance is $D$$\approx$40 Mpc,
which is in agreement with the distance modulus calculated in \citet{miller2020_19yvq}.
Both to be consistent with \citet{siebert_19yvq} and \citet{tucker_19yvq},
and because using the SBF distance value would further decrease the object's already low luminosity,
we adopt the distance modulus from \citet{miller2020_19yvq} throughout this work ($\mu = 33.14 \pm 0.11$, $D=42.5\pm2.2$ Mpc).
We also adopt a Milky Way extinction value of $E(B-V)$=0.017 mag using the \citet{Schlafly_2011} calibration of the \citet{schlegel98} dust maps.

\begin{figure*}
\begin{center}
\includegraphics[width=\textwidth]{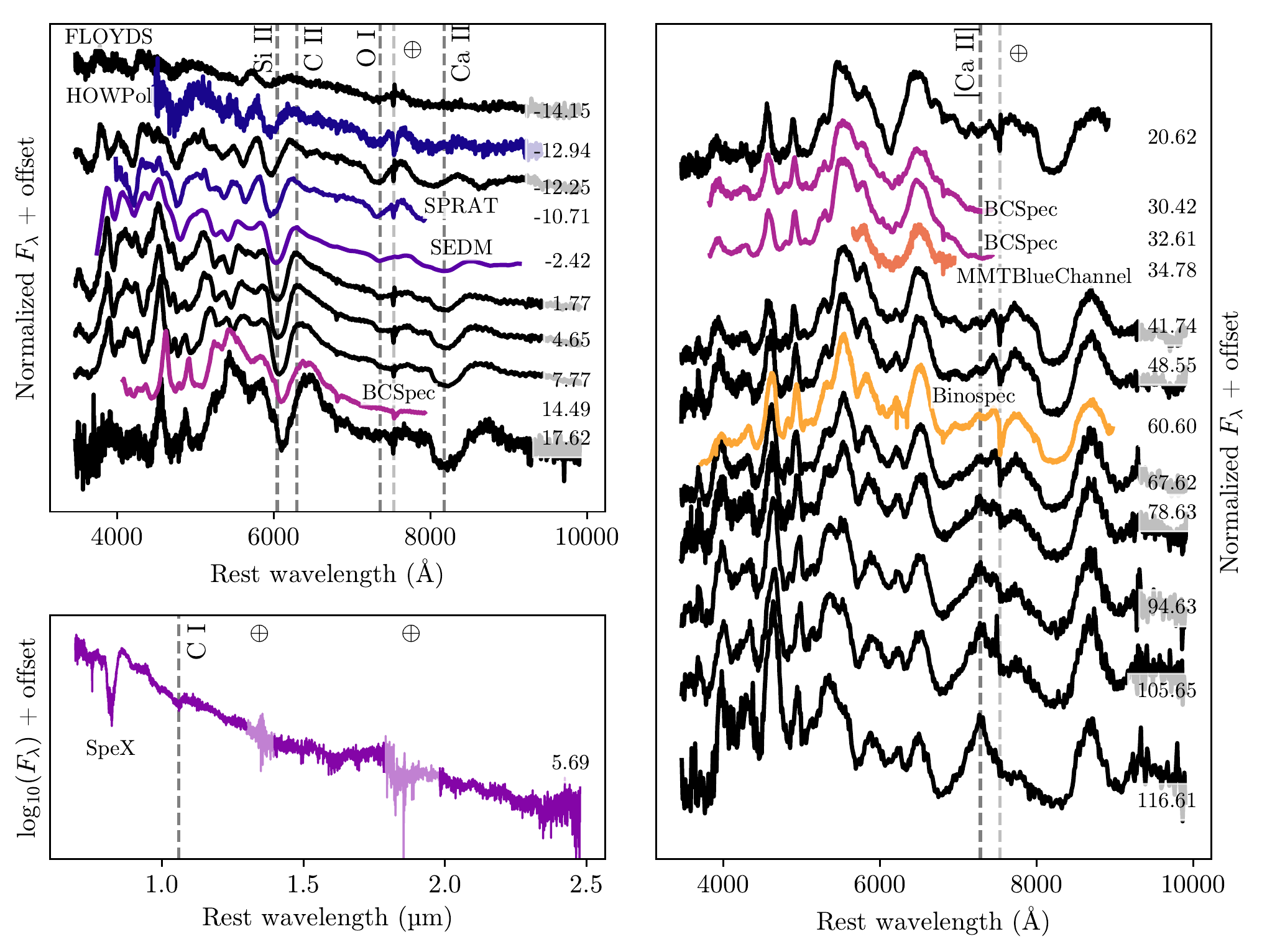}
\caption{
The top left and right-hand panels indicate the optical spectral evolution of SN 2019yvq,
separated into panels purely for readability.
The bottom left panel shows the IR spectrum at $\sim$6 days taken with SpeX on the IRTF (Section \ref{ssec:obs_spec}).
Epochs (in days) with respect to \textit{B}-band maximum are included as labels on each spectrum.
The wavelengths of spectral features are marked with dashed lines,
corresponding to their approximate velocity which they have at maximum light to guide the eye in tracking their velocity evolution.
Telluric features are marked with $\oplus$.
The primary source for spectra was the FLOYDS instrument at Las Cumbres (black spectra),
but a number of other spectra (detailed in Sections \ref{sec:discovery} and \ref{ssec:obs_spec})
are included as well.
The final three spectra have been binned by a factor of 5, for clarity.
\label{fig:spec_timeseries}
}
\end{center}
\end{figure*}

\subsection{Photometry} \label{ssec:obs_phot}

Figure \ref{fig:lc} displays our full photometric dataset.

An intense \textit{UBVgri} follow-up campaign was undertaken using the 1-m telescopes of Las Cumbres Observatory \citep[LCO;][]{LCO}.
Data were reduced using \texttt{lcogtsnpipe} \citep{Valenti_2016} by performing PSF-fitting photometry.
Zeropoints for images in the \textit{UBV} filters were calculated from Landolt standard fields \citep{landolt} taken on the same night by the same telescope.
Likewise, zeropoints for images in the \textit{gri} filter set were calculated by using Sloan magnitudes of stars in the same field as the object \citep{SDSS}.

Observations from the {\it Neil Gehrels Swift Observatory} \citep[{\it Swift};][]{Gehrels04} and the Ultra-Violet Optical Telescope \citep[UVOT;][]{Roming05} were obtained
under GI Program 1518168
and reduced using the pipeline associated with the {\it Swift} Optical Ultraviolet Supernovae Archive \citep[SOUSA;][]{Brown_etal_2014_SOUSA} and the zeropoints of \citet{Breeveld10}.
The temporal sensitivity changes were corrected for using the 20200925 CALDB\footnote{\url{https://heasarc.gsfc.nasa.gov/docs/heasarc/caldb/swift/docs/uvot/uvotcaldb_throughput_06.pdf}}.  
Template observations from 2012 were used to subtract the host galaxy count rates from the \textit{UVW2}, \textit{UVM2}, and \textit{UVW1} filters.

In addition to the Las Cumbres and {\it Swift} photometric data, we have also obtained unfiltered photometry taken with the Itagaki Astronomical Observatory's Celestron 14-inch telescope in the days after discovery, including the nondetection taken the day prior to SN 2019yvq's discovery.

We gather $g$ and $r$ band data from the public ZTF data stream using the \texttt{MARS} transient broker\footnote{\url{https://mars.lco.global/}},
and present the near-peak data in Figure \ref{fig:lc} as comparison.



\subsection{Spectroscopy}\label{ssec:obs_spec}

Figure \ref{fig:spec_timeseries} displays our full spectroscopic dataset.


A sequence of optical spectra were taken primarily with the FLOYDS spectrograph mounted on Las Cumbres Observatory's 2-m telescope on Haleakala, HI, and were reduced as described in \citet{Valenti14}.

Additional optical spectroscopy was obtained with the 2.3-m Bok telescope and the B\&C spectrograph using both the 300 line/mm grating and a higher resolution 1200/mm line grating.
We also obtained an MMT medium resolution (1200 l/mm) spectrum on  2020-02-18 11:27 UTC using the Blue Channel spectrograph \citep{Schmidt89}.
These data were reduced using standard IRAF tasks.
We use the Na ID doublet in the high resolution data as one method of estimating host galaxy extinction from cold gas as discussed in Section \ref{ssec:Na_ID}.

Finally, a near-infrared spectrum of SN~2019yvq was taken on 2020 Jan 20 (UT) with SpeX \citep{Rayner03} on the NASA Infrared Telescope Facility in cross-dispersed `SXD' mode, providing wavelength coverage from $\sim$0.8--2.4 $\mu$m; these data were reduced in a standard way, as described in \citet{Hsiao19}.

All new data are made publicly available on the Weizmann Interactive Supernova Repository\footnote{\url{https://wiserep.weizmann.ac.il/}}\citep{Yaron12}.

\section{Data Analysis} \label{sec:data_analysis}

\subsection{Lightcurve and Color Evolution Analysis} \label{ssec:lightcurve_analysis}


\begin{table}[t!]
\begin{center}
\begin{tabular}{ |c|c|c|c| } 
 \hline
 Method & $E(B-V)$ & $\sigma_{E(B-V)}$ & $M_B$ \\ 
 \hline
 Na ID                    & 0.052 & $_{-0.025}^{+0.053}$ & -18.41 \\ 
 Lira Law                & 0.268 & 0.043 & -19.29 \\ 
 \texttt{SNooPy}          & 0.342 & $0.031 \pm 0.060 \textrm{\ (sys)}$ & -19.60 \\ 
 \texttt{SNooPy} (no $i$) & 0.445 & $0.049 \pm 0.060 \textrm{\ (sys)}$ & -20.02 \\ 
 SALT2                    & 0.347 & 0.015 &  -19.62 \\ 
 SALT2 (no $i$)           & 0.631 & 0.019 &  -20.78 \\ 
 MLCS2k2                  & 0.252 & 0.0036 &  -19.23 \\ 
 MLCS2k2 (no $i$)         & 0.279 & 0.0038 &  -19.34 \\ 
 \hline
\end{tabular}
\caption{Range of extinction values and peak absolute magnitudes computed using different methods and SN Ia fitting programs.
SALT2 and MLCS2k2 fits were done using the \texttt{sncosmo} package
and Lira Law fits were done with a fixed slope,
as discussed in the text.
We adopt the Na ID extinction value throughout our analysis.
}
\label{table:extinction}
\end{center}
\end{table}

The lightcurve of SN 2019yvq is presented in Figure \ref{fig:lc}.
The most striking feature of this lightcurve is the strong wavelength-dependent excess of the first epoch,
seen in data from Las Cumbres, ZTF, and \textit{Swift}.
We note especially the excess in the mid-UV \textit{Swift} filters,
where the magnitude during the initial bump is brighter than the ``peak" magnitude.
This is even more extreme than other objects with an observed mid-UV excess at early times such as SN 2012cg \citep{Marion16} and iPTF14atg \citep{Cao15}. 
We also note that SN 2017cbv \citep{Hosseinzadeh17},
the SN Ia with the most clearly resolved early optical blue bump,
displayed only a moderate excess in the 
\textit{UVW1},
\textit{UVM2}, or
\textit{UVW2}
bands compared to what is expected from companion shock interaction models (as shown in Figure 3 of that paper),
although its UV colors are still quite blue compared to other normal SNe Ia \citep{brown_red}.

Different methods of estimating the extinction due to the host galaxy of SN 2019yvq yielded significantly different results,
as 
summarized
in Table \ref{table:extinction}.
For all fits we fixed $R_{V,\textrm{host}} = 3.1$.

\begin{figure}[t]
\includegraphics[width=0.49\textwidth]{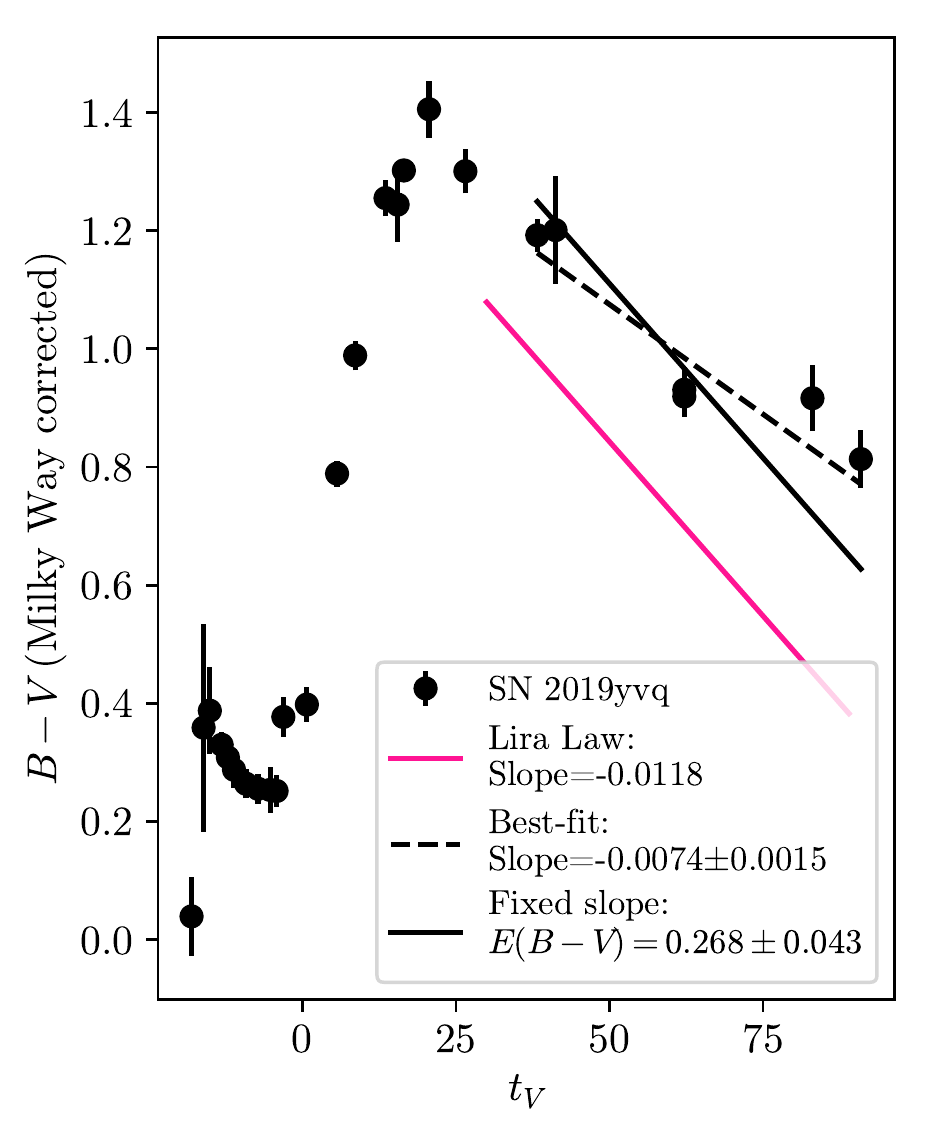}
\caption{ 
Comparisons of the $B-V$ color evolution of SN 2019yvq (black) to the Lira Law (pink).
The best-fit line (dashed) to the appropriate SN 2019yvq data has a slope $2.9\sigma$ away from the expected slope.
Fixing the slope (solid line) is one method of measuring the host extinction,
reported in Table \ref{table:extinction}.
Following the convention of \citet{Phillips99},
data are plotted relative to $t_V$ (days from \textit{V}-band maximum).
\label{fig:lira_law}}
\end{figure}

\begin{figure*}
\begin{center}
\includegraphics[width=\textwidth]{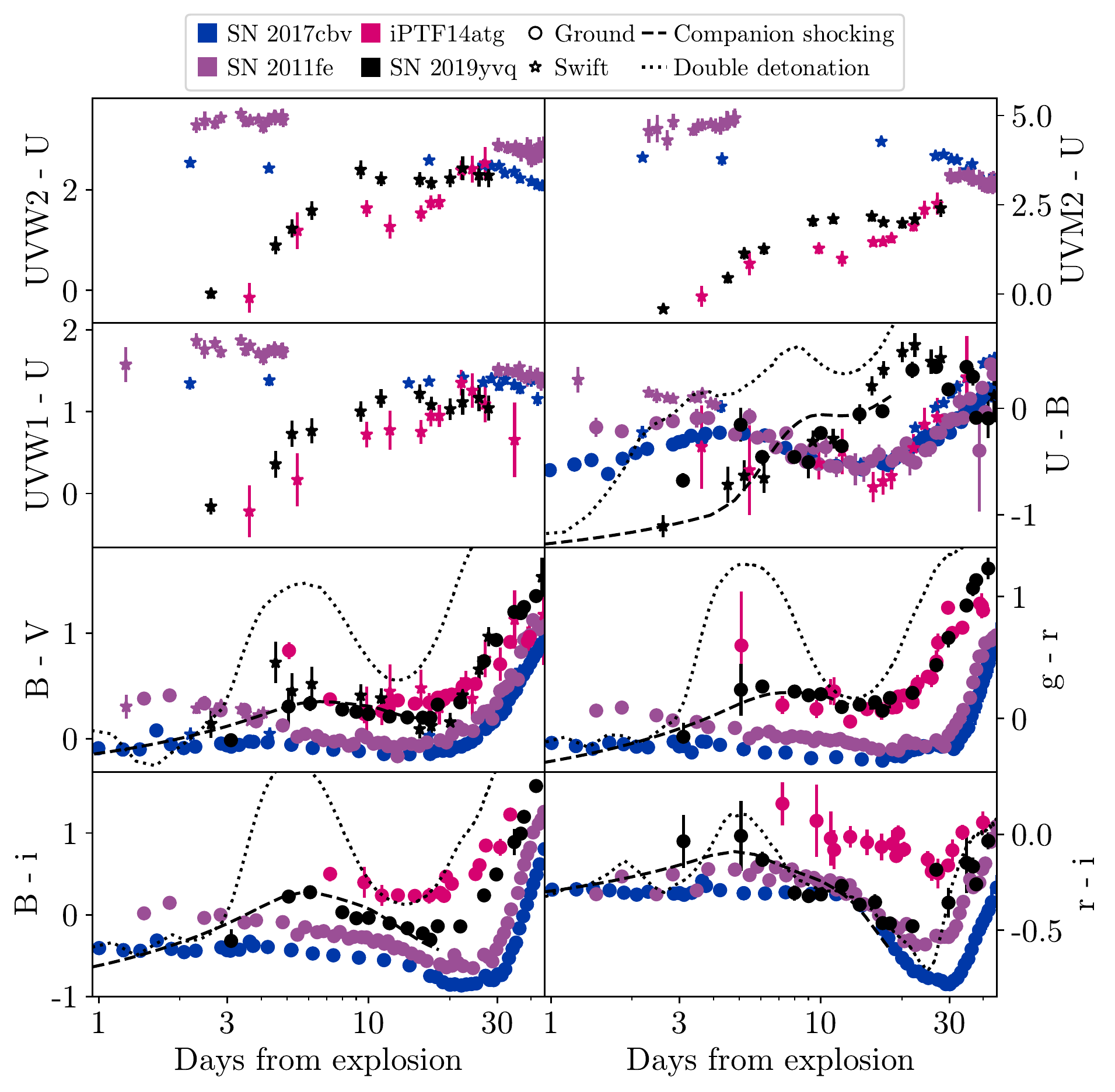}
\caption{Color evolution of SN 2019yvq compared with other SNe Ia.
We assume an explosion epoch of SN 2019yvq derived from the best-fit companion shocking model,
and the two sets of model colors plotted are the best-fit models described in Section \ref{sec:models}.
We note again the extremely strong early blue color in every filter combination besides $r-i$.
\label{fig:color}}
\end{center}
\end{figure*}

One method of calculating extinction in SNe Ia is the ``Lira Law."
As shown in Figure 1 of \citet{Phillips99},
the $B-V$ color evolution of many SNe Ia is similar between 30 and 90 days after $V$ maximum,
and can be fit with a line described by Equation 1 of that paper.
That expected linear color evolution is shown in pink in Figure \ref{fig:lira_law}.
$E(B-V)$ can then be measured by fitting a line with the same slope to the color data,
and finding the linear offset needed to deredden the fit to the expected Lira Law values.
Using this method we measure $E(B-V)=0.268\pm0.043$ for SN 2019yvq.
However, the $B-V$ color evolution of SN 2019yvq has a best-fit slope $2.9\sigma$ away from the slope predicted by the Lira Law.
The shallower slope of SN 2019yvq is not unprecedented \citep[see e.g.][]{forster_liralaw},
but does cast doubt on the $E(B-V)$ value obtained from the Lira Law comparison.

We also attempted to fit the \textit{BVgri} data from Las Cumbres using the \texttt{SNooPy} software package \citep{snpy}.
We obtained the extinction value by comparing to \texttt{EBV\_model},
which required a high extinction value (0.342) to match the data.
similar to the findings in \citet{miller2020_19yvq}.
The fits start at a phase of -10 days with respect to maximum light, and thus the early excess should not bias the results.
We found that the fits strongly overpredicted the secondary \textit{i} maximum,
so we also performed fits which excluded those data.

In contrast to normal SNe Ia,
SN 2019yvq lacks a strong secondary NIR peak,
although \citet{tucker_19yvq} do find evidence of a weak secondary NIR maximum in both the ZTF \textit{i}-band data and the TESS lightcurve.
We take this very weak secondary NIR peak as one of several pieces of evidence that the object is intrinsically underluminous compared to normal SNe Ia (see Section \ref{sec:02es_comp}).

We repeated this process on the $UBVgri$ Las Cumbres data using the SALT2 \citep{salt2} and MLCS2k2 \citep{mlcs2k2} fitting packages,
accessed through SNCosmo \citep{sncosmo} with an added \texttt{CCM89Dust} component to measure $E(B-V)$.
We exclude the first three epochs of data,
to reduce biases from attempting to fit the early blue excess.
The fits were generally poor: in order to achieve a $\chi_{\textrm{reduced}}^2$ of less than 2 on the best fits (MLCS2k2, no \textit{i} band),
we required a systematic error of more than three times the average flux error to be added in quadrature at each point.
In general the fits again overpredicted the secondary \textit{i}-band peak.
Values for the \texttt{SNooPy} and SNCosmo fits are reported in Table \ref{table:extinction}.

The fact that different methods of estimating $E(B-V)$ led to such a wide range of extinction values,
and the fact that methods which relied on fitting to SN Ia templates resulted in generally poor fits,
led us to conclude that SN 2019yvq is an inherently peculiar SN Ia.
We therefore
adopt the extinction value obtained from fitting the Na ID lines, $E(B-V)=0.052^{+0.053}_{-0.025}$ (see Section \ref{ssec:Na_ID} for methodology).
This value, while significantly lower than other possible values,
results in an underluminous peak absolute magnitude,
which is consistent with SN 2019yvq's weak secondary IR maximum and high lightcurve decline rate.
Additionally, it is consistent with the value calculated in \citet{miller2020_19yvq}
($E(B-V)_{\textrm{host}}\approx0.032$),
which they derive using the same method, but a different spectrum.
\citet{siebert_19yvq} and \citet{tucker_19yvq} adopt this value from \citet{miller2020_19yvq},
so our extinction value is also consistent with all previously published work on SN 2019yvq.

We fit a fifth-order polynomial to the near-peak \textit{B} data to obtain standard lightcurve parameters.
We find that SN 2019yvq reached its peak apparent magnitude of
$B_{\textrm{max}} = 15.01 \pm 0.03$
($M_B = -18.4 \pm 0.1$)
on MJD
$58862.8 \pm 0.4$,
with
$\Delta m_{15}(B) = 1.36 \pm 0.10$.
We note that this $\Delta m_{15}$ is lower than the value inferred by \citet{miller2020_19yvq} from the \textit{g} lightcurve and used in \citet{siebert_19yvq}.

The color evolution of SN 2019yvq is presented in Figure \ref{fig:color}.
The \textit{Swift} data for all objects were extinction-corrected using the method of \citet{brown_extinction} (Table 1).
We note that SN 2019yvq becomes rapidly redder in all optical colors (besides $r-i$) over the first five days.
In $(B-V)$ and $(g-r)$ especially, it is much redder than typical SNe Ia
such as SN 2011fe
\citep[data from][]{zhang_11fe}
and more closely mirrors the evolution of iPTF14atg.
iPTF14atg was also an underluminous SN Ia with a strong early UV excess \citep{Cao15},
and belonged to the 02es-like subclass, whose namesake is described in \citet{Ganeshalingam12}.
As discussed in Section \ref{sec:02es_comp}, we classify SN 2019yvq as a transitional 02es-like.

In terms of {\it Swift} UV colors, SN 2019yvq stands out even more compared to typical SNe Ia,
and is $\gtrsim$1 magnitude bluer than SN 2017cbv in $(UVW1-U)$ at $\sim$5 days after the estimated explosion time.
This extreme UV color and subsequent evolution is again most similar to iPTF14atg within ten days of explosion.

\begin{figure*}
\begin{center}
\includegraphics[height=2.2in]{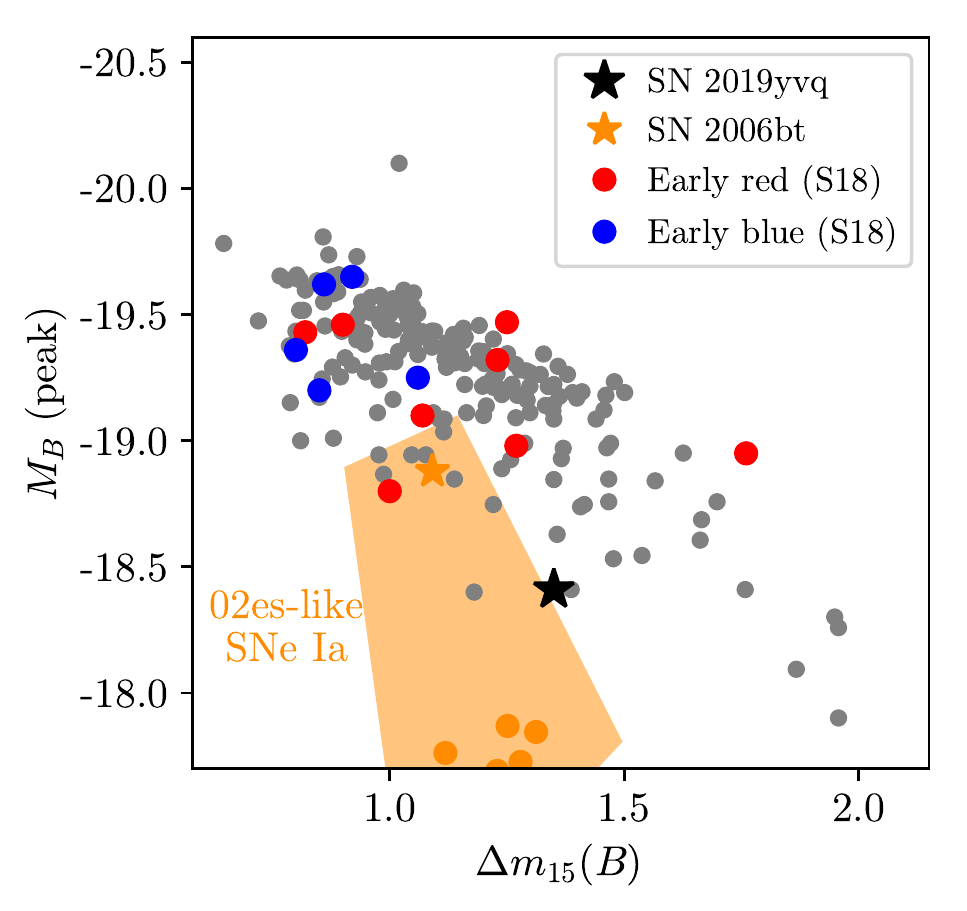}
\includegraphics[height=2.2in]{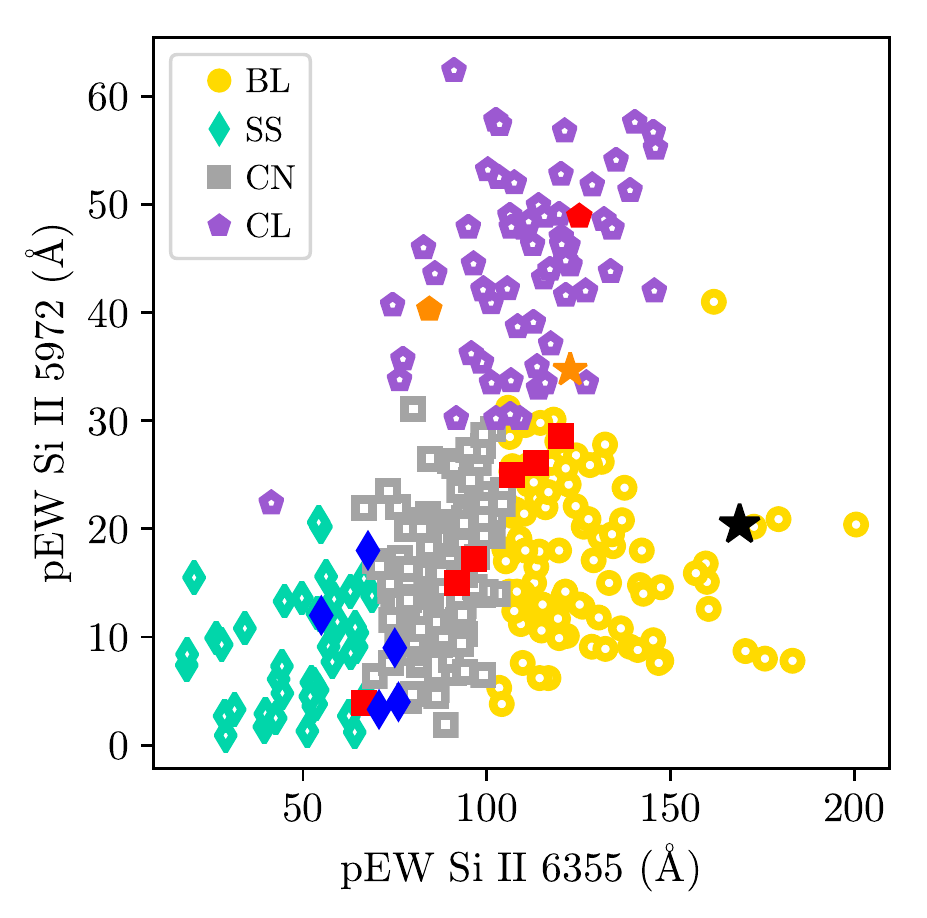}
\includegraphics[height=2.2in]{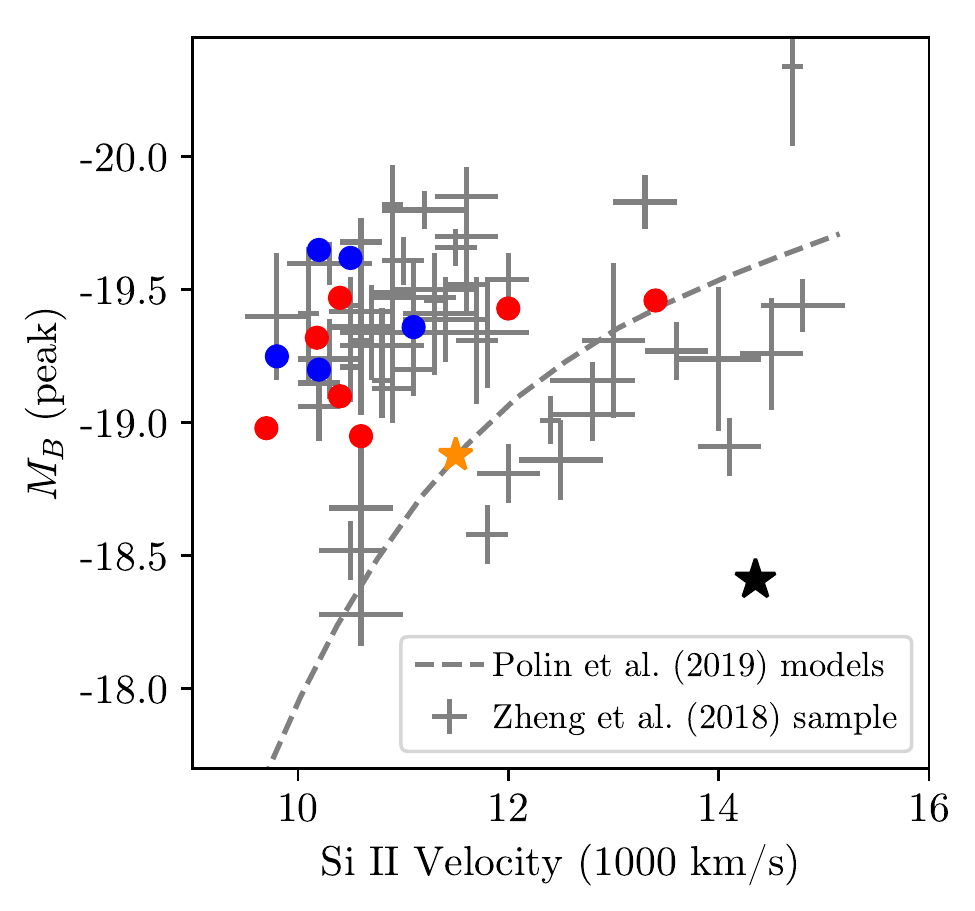}
\caption{ Demographic properties of SN 2019yvq (black star in each plot).
We note that SN 2019yvq is at the edge of normal parameter space in several respects,
and is well-separated from the early blue objects of S18.
It is instead closer to (although still substantially different from) the transitional 02es-like SN 2006bt (orange star in each plot).
\textit{Left}: Luminosity decline rate relation for SNe Ia,
with the gray background points coming from the union of samples presented by several groups 
\citep{Blondin12,Folatelli12}.
The orange polygon and data points replicate the sample of 02es-like SNe Ia in \citet{taubenberger_SN_handbook},
with the transitional SN 2006bt represented by the orange star in each plot.
In blue and red we show the early SN Ia sample presented by S18,
split by their early light curve colors.
Out of the S18 sample,
we have adjusted the absolute magnitude of SN 2017cbv to match the distance of 
$D=12.3$ Mpc found in \citet{Sand18}.
\textit{Center}: The location of SN 2019yvq (black star) in the Branch diagram \citep{Branch06},
which groups SNe Ia as broad line (BL), shallow silicon (SS), core normal (CN), or cool (CL) based on the pseudo-equivalent widths of two Si II features.
The background sample is the same as the left panel,
and the only other 02es-like (in orange) in \citet{Blondin12} is SN 2002es itself.
\textit{Right}: A replica of the plot from \citet{polin_subch}
comparing $0.01$ \msun \ He shell double detonation models
to a sample of SNe Ia from \citet{Zheng2018},
with velocities measured at peak.
The prototype object SN 2002es has a Si II velocity which is too low (5890 km s$^{-1}$) to fit in the axis range of these plots.
\label{fig:spec_demographics}}
\end{center}
\end{figure*}

Based on the lightcurve parameters, we can begin to put SN 2019yvq in context with other SNe Ia,  especially those with early light curve data as well.
In the left panel of Figure~\ref{fig:spec_demographics}, we show the $M_B$ versus $\Delta m_{15}(B)$ relation of \citet{Phillips93}, populated with a large sample of nearby SNe Ia \citep[see Figure 14 from \citealt{Parrent14}, with original data from ][]{Blondin12,Folatelli12,Pakmor13}.
When we include the ``blue" and ``red" sample of early SN Ia of \citet{redvsblue} (hereafter S18), we see the tendency of early blue objects to be slower declining and slightly brighter than the red sample.
SN 2019yvq notably stands out from the ``early-blue" sample with its much higher decline-rate.
In this parameter space it is closer to another transitional 02es-like,
SN 2006bt (the orange star in Figure \ref{fig:spec_demographics}),
although still well-separated from that object.

\subsection{Spectral Analysis} \label{ssec:spec_analysis}

We show the spectral evolution of SN 2019yvq in Figure~\ref{fig:spec_timeseries},
from roughly $-14$ to $+117$ days with respect to $B$-band maximum. Using the Supernova IDentification software package \citep[\texttt{SNID};][]{snid} on the FLOYDS spectrum taken at $+$1.8 d with respect to $B$-band maximum we find that all reasonable matches correspond to normal SN Ia.
In particular, the spectrum is well matched to SN~2002bo near maximum light 
except in the region of $\sim$4000--4500 \AA,
which we 
attribute to weak Ti II absorption
and 
discuss further in Section \ref{sec:02es_comp}.
We note that the initial spectrum of SN~2019yvq shows faint H$\beta$, H$\alpha$ and [N II] emission; upon investigation, we believe this emission is from the host galaxy due to slight mis-centering of the SN within the slit.

\subsubsection{Velocities and Spectral Classification}

We measure a Si II $\lambda$6355 velocity of 14,400 km s$^{-1}$ near maximum light, as well as pseudo-equivalent width (peW) values of 169 \AA~and 20 \AA~for the Si II $\lambda$6355 and $\lambda$5972 features, respectively, from the +1.8d FLOYDS spectrum (these measurements, and those that follow, are in broad agreement with those of \citealt{miller2020_19yvq}).
Here SN~2019yvq is clearly a high-velocity (HV) object in the \citet{Wang09} classification scheme (e.g. objects with Si II $\lambda$6355 $\gtrsim$11,800 km s$^{-1}$ near max).
To put SN~2019yvq in the context of the standard Branch classification scheme \citep{Branch06}, we plot it along with a larger sample of SNe Ia \citep{Blondin12} in the center panel of Figure~\ref{fig:spec_demographics}.  Here SN~2019yvq is clearly a Broad Lined (BL) SN Ia, with a very deep and broad Si II $\lambda$6355 feature.  This is consistent with its match to SN~2002bo, which was another BL event.  
We also plot the blue and red sample from S18 on the Branch diagram, and note that SN~2019yvq again stands alone among the early blue objects as a broad lined event,
as most of the others are Shallow Silicon or Core Normals,
and instead it is closer to the transitional 02es-like SN 2006bt.


To explore the demographic place of SN~2019yvq further, we plot the Si II $\lambda$6355 velocity near maximum light versus the absolute $B$-band magnitude in the right panel of Figure~\ref{fig:spec_demographics}.  This plot is largely a reproduction of Figure~11 in \citet{polin_subch}, with the grey data points originating from the SNe Ia sample of \citet{Zheng2018}; the blue and red sample of S18
and SN 2006bt
are plotted as well.  As discussed by \citet{polin_subch}, two groups of SNe Ia are apparent in the plot: one that is tightly clumped at $v\approx10,500$ km s$^{-1}$ and $M_B\approx-19.4$ and is attributed to Chandrasekhar mass explosions, and a second group that follows a relationship between luminosity and velocity, roughly tracking expectations from the sub-Chandrasekhar class of explosions, as illustrated by the dashed line which depicts a set of 0.01 $M_{\odot}$ He shell double detonation models.  
It is clear that SN~2019yvq is not well-matched by either population, and a model with different He shell mass is needed to replicate its position,
as is found in Section \ref{ssec:double_det}.


\subsubsection{Search for Unburned Carbon}

The presence of unburned carbon in SN Ia spectra is potentially a powerful discriminant between explosion models.  Chandrasekhar-mass delayed detonation explosions predict complete carbon burning for normal-bright SNe Ia \citep[e.g.][]{Kasen09}, and increasing amounts of unburned carbon for fainter SNe Ia \citep[e.g.][]{Hoflich02}.   In the explosions of sub-Chandrasekhar mass white dwarfs, on the other hand, the initial surface detonation may leave little or no detectable carbon \citep[e.g.][]{Fink10,polin_subch}.

The most commonly searched for carbon feature is C~II $\lambda$6580\AA, which can be difficult to detect both because it fades quickly after explosion and is near the strong Si II $\lambda$6355\AA~absorption line.  Large spectroscopic samples have found that $\sim$20-30\% of early time SNe Ia data have C~II signatures, with the chances of detection increasing the earlier the data were taken \citep{Thomas11,Parrent11,Folatelli12,Silverman_carbon,wyatt_carbon}.  Interestingly, several of the SN Ia with early light curve excesses have also displayed strong early carbon, including SN~2017cbv \citep{Hosseinzadeh17}, iPTF16abc \citep{Miller18} and SN2018oh \citep{li_18oh}.

We have closely inspected all of our SN~2019yvq optical spectra through maximum light at the expected position of C II $\lambda$6580 \AA, near the red shoulder of the Si II $\lambda$6355 \AA~absorption line.  No C II feature is apparent, and our earliest data do not show the strong carbon absorption seen in SN~2017cbv and iPTF16abc, although the signal to noise of our early data is not good enough to make definitive claims on any weak C II feature.  We have further inspected our IRTF spectrum taken at +6~d with respect to $B$-band maximum, as it has been suggested that the C I $\lambda$1.0693 $\mu$m line is a good tracer of unburned carbon.
No C I line is apparent, but this spectrum is later than ideal since this feature is most visible around maximum light \citep[e.g.][]{Hsiao13,Hsiao19}.
Detailed modeling is necessary to completely rule out any subtle carbon feature, but this is beyond the scope of the current work.  

In conclusion, we can make no definitive claim about the presence of either C II $\lambda$6580 \AA~or C I $\lambda$1.0693 $\mu$m, partially due to low signal to noise data, although we can rule out the strong carbon seen in previous SNe Ia with blue light curve excesses.   This lack of strong carbon is in broad agreement with expectations from sub-Chandrasekhar helium shell detonation models \citep[e.g.][]{polin_subch}, which we explore further in our model comparisons below.

\subsubsection{Medium Resolution Spectra and Na ID}
\label{ssec:Na_ID}
The Na ID doublet is often used to estimate host galaxy extinction in nearby SNe \citep[e.g.][]{Poznanski12}, although the correlation between host extinction and Na ID equivalent width has a large scatter
\citep[e.g.][]{galbany_NaID}.
Although the diffuse interstellar band at 5780\AA~ has been shown to be a superior tracer of host extinction \citep{Phillips13}, we do not detect the line in our medium resolution Bok spectrum. 
The Na ID doublet at the redshift of SN2019yvq's host ($z$=0.00908) is clearly visible in our medium resolution Bok B\&C spectrum ($R$$\approx$3400) taken on 2020 January 29 UT (a medium resolution MMT Blue Channel spectrum taken on 2020 February 18 does not have sufficient signal to detect the doublet), and we measure 0.28\AA~ and 0.18\AA~ for the equivalent width of the D1 and D2 lines, respectively.  Using the correlation found by \citet{Poznanski12}, this translates to an expected host extinction of 
$E(B-V)_{\textrm{host, Na ID}}=0.052^{+0.053}_{-0.025}$ mag.
As discussed in Section \ref{ssec:lightcurve_analysis}, this is the host extinction value we use throughout the paper.

\begin{figure}
\begin{center}
\includegraphics[width=0.5\textwidth]{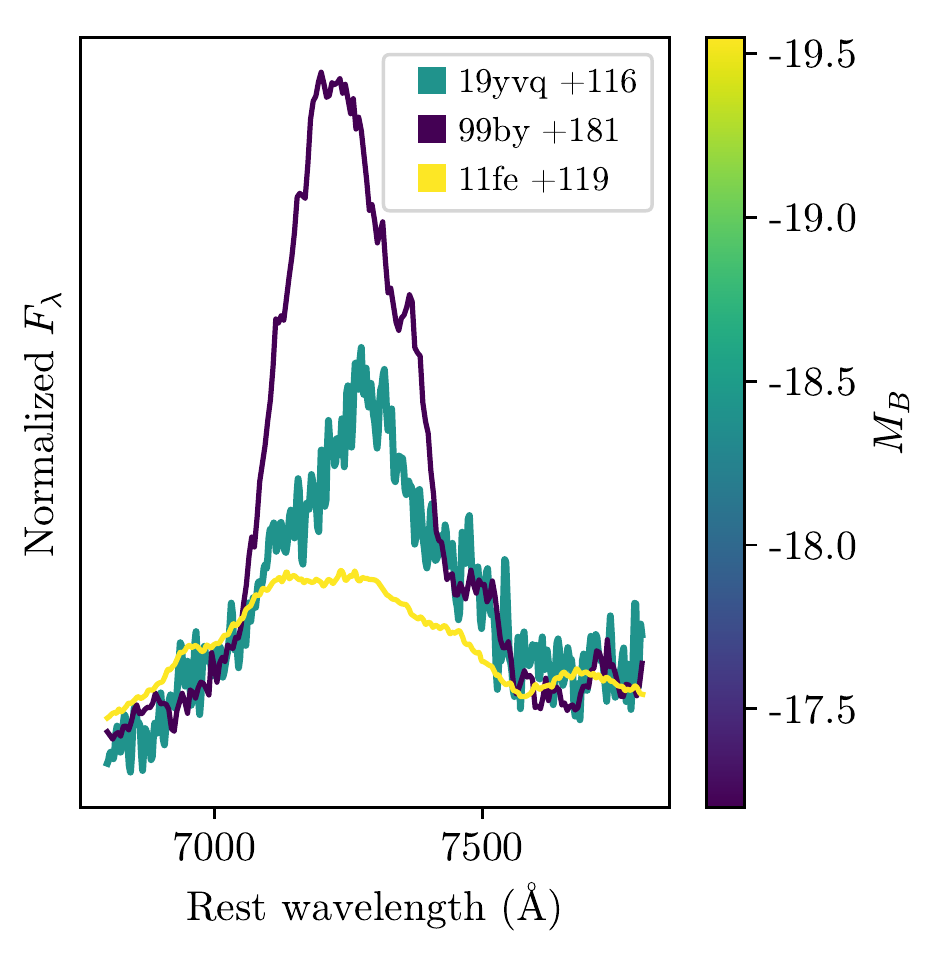}
\caption{
Nebular spectra of SNe Ia focusing on the [Ca II], [Fe II], [Ni II] line complex.
This feature is strongest in the nebular spectra of underluminous SNe Ia,
and is the subject of thorough modeling in \citet{siebert_19yvq} for a +153d Keck spectrum of SN 2019yvq.
The legend displays the shortened SN name (e.g. SN2019yvq $\rightarrow$ 19yvq) and the epoch in days after $B$ maximum.
Spectra have been normalized to have identical mean fluxes over their full wavelength range ($\sim$3500--10000 \AA).
SN 2019yvq lies in between normal SNe Ia (represented by SN 2011fe) and low-luminosity SNe Ia (represented by the 91bg-like SN 1999by).
}
\end{center}
\label{fig:nebular_spec}
\end{figure}

\subsubsection{Nebular spectra of SN 2019yvq}
\label{ssec:nebular_spec}

The nebular spectra of SNe Ia can provide an independent way to differentiate between progenitor systems,
since different progenitors and explosion channels should have different nebular signatures.

The violent merger of two WDs should result in nebular [O I] due to its ejection at low velocities \citep{pakmor_violentmerger},
although this has only been seen in the nebular spectra of the 02es-like SN 2010lp \citep{taubenberger_10lp} and is not present in the nebular spectra of SN 2019yvq.

The double-detonation scenario should only partially burn the core,
leaving strong Ca signatures \citep{polin_nebularca}.
SN 2019yvq does display nebular [Ca II] which is intermediate in strength between typical- and low-luminosity SNe Ia,
as shown in Figure \ref{fig:nebular_spec}.

Lastly, the companion interaction scenario should produce H and He emission from the swept-up material \citep{Boty18,dessart_2020},
although this is seen in an extremely limited number of cases \citep{kollmeier_nebularHa, prieto_20}.
We use the nebular spectra of SN 2019yvq to measure limits on the luminosity and mass of swept-up H and He, following the methodology of \citet{sand20} and references therein. To briefly summarize, we first smooth the spectrum on a scale much larger than the expected width of an H$\alpha$ feature.
We then subtract off the smoothed spectrum and search for any excess flux in the residuals, assuming an expected width of FWHM $\approx$ 1000 km s$^{-1}$ (22~\AA) for the line width and a potential offset from the rest wavelength of up to $\sim$1000 km s$^{-1}$ as well.
Following Equation 1 from \citet{Boty18}, we then estimate the mass of the stripped material, 
after predicting the luminosity of SN 2019yvq at +200 days.
For the nebular spectrum taken +106 days past maximum, 
$M_\textrm{H} < 1.6\times10^{-3} \textrm{ \msun}$
and 
$M_\textrm{He} < 2.0\times10^{-2} \textrm{ \msun}$ (using the He~I~$\lambda$6678 line).
Using an additional nebular spectrum taken +117 days past maximum, 
$M_\textrm{H} < 1.7\times10^{-3} \textrm{ \msun}$
and 
$M_\textrm{He} < 2.1\times10^{-2} \textrm{ \msun}$.
With access to a higher signal-to-noise spectrum,
\citet{siebert_19yvq} place even stricter limits on the amount of swept-up He and He:
$M_\textrm{H} < 2.8\times10^{-4} \textrm{ \msun}$
and
$M_\textrm{He} < 2.4\times10^{-4} \textrm{ \msun}$.

\begin{table}[t!]
\begin{center}
\begin{tabular}{ |c|c|c| } 
 \hline
 Parameter & 02es-like SNe Ia & SN 2019yvq \\
 \hline
 $M_B$ & -17.6 -- -18.1 & -18.41 \\
 $\Delta m_{15}(B)$ & 1.1 -- 1.3 & 1.36 \\
 Rise time (days) & 19 -- 20 & 18.7 \\
 $(B-V)_{\textrm{max}}$ & 0.2 -- 0.5 & 0.22 \\
 Secondary IR maximum & Weak & Weak \\
 $v_{\textrm{Si II}}$ (km s$^{-1}$) & 6000 -- 10000 & 14400 \\
 Ti II at peak & Yes & Yes \\
 nebular [Fe II] and [Ca II] & Yes & Yes \\
 \hline
\end{tabular}
\caption{Comparisons between SN 2019yvq and 02es-like SNe Ia.
Parameter ranges for 02es-like SNe Ia are taken from \citet{taubenberger_SN_handbook} and are intended to be approximate, reflecting the small sample size and diversity of this subclass.
}
\label{table:02es}
\end{center}
\end{table}

The combination of the presence of [Ca II] and a lack of narrow hydrogen emission is consistent with a double-detonation progenitor system,
which is what is inferred by \citet{siebert_19yvq}.
Despite these limits,
we cannot unequivocally claim that SN 2019yvq is a double detonation event
due to discrepancies in best-fit models of photospheric photometry and nebular spectroscopy.
Our conclusion in this regard is in agreement with \citet{tucker_19yvq} and \citet{miller2020_19yvq},
and is discussed in more detail in Section \ref{ssec:double_det}.

\begin{figure*}[t!]
\begin{center}
\includegraphics[width=\textwidth]{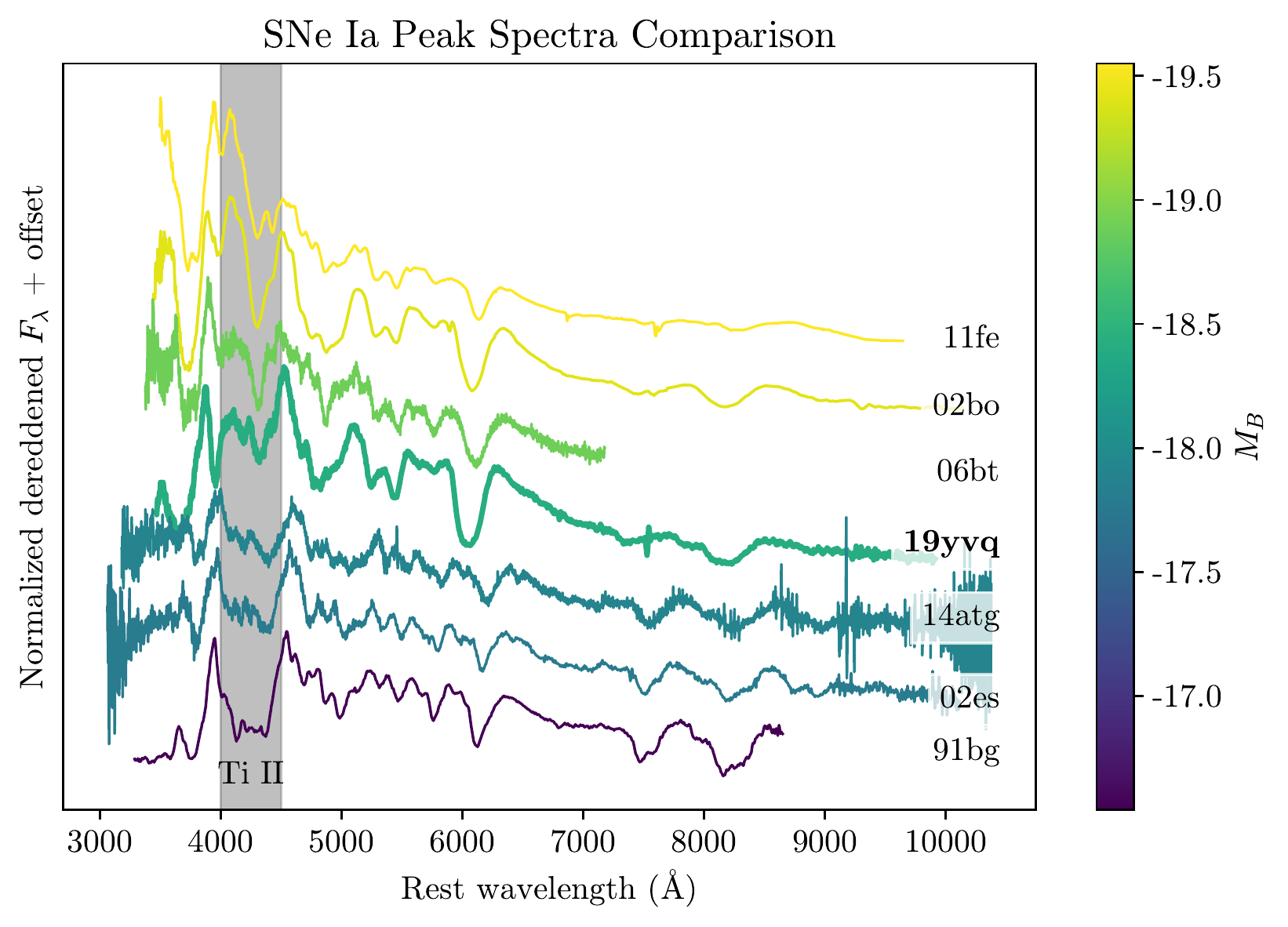}
\caption{Comparisons of SNe Ia peak spectra over a wide range of luminosities.
Although the spectrum of SN 2019yvq is quite similar to SN 2002bo (a more typical luminosity SN Ia),
its primary difference is in the $\sim$4000--4500 \AA \ region.
This coincides with the ``titanium trough" present in lower luminosity SNe Ia,
and SN 2019yvq's extra absorption in this wavelength region supports the interpretation of it as an underluminous SN Ia despite obvious differences when comparing to the spectrum of SN 2002es.
The combination of low temperature and luminosity with broad high-velocity Si II is rarely seen in SNe Ia and is difficult to reproduce in models.
}
\label{fig:spec_comp_L_range}
\end{center}
\end{figure*}

\section{Comparisons to SN 2002es}\label{sec:02es_comp}

SN 2019yvq shares some characteristics with 02es-like SNe Ia, and could be considered an 02es-like depending on how broad a definition of that subclass is taken.
We classify it as a transitional 02es-like.
Although this term has not previously been used in the literature to describe any objects,
it accurately reflects the nature of SN 2019yvq.
Table \ref{table:02es} summarizes various photometric and spectroscopic signatures of 02es-like SNe Ia,
taken from \citet{taubenberger_SN_handbook}.
See \citet{Ganeshalingam12} for a study of the eponymous SN 2002es, and \citet{taubenberger_SN_handbook} and \citet{white15} for reviews of this subclass.

SN 2019yvq is at the edge of what could be considered 02es-like in several respects.
Its peak brightness and lightcurve width are on the edge of the class,
as seen in the left panel of Figure \ref{fig:spec_demographics}.
Like 02es-like SNe Ia,
SN 2019yvq also displays an almost nonexistent secondary IR maximum and red colors after its initial blue excess
(see Figure \ref{fig:color} and its similarity to the 02es-like iPTF14atg).

Spectroscopically there are both similarities and obvious differences,
as highlighted in Figure \ref{fig:spec_comp_L_range}.
The peak spectrum of SN 2019yvq is most similar to SN 2002bo,
which also displayed deep Si II 6355 and had a similar Si II line ratio.
SN 2002bo had a more typical peak luminosity for SNe Ia \citep[$M_B=-19.41$,][]{benetti_02bo}.
SN 2019yvq's Si II velocity and line ratio
make it an outlier compared to other 02es-like SNe Ia,
since these spectral features would normally indicate an energetic and luminous event.
Figure \ref{fig:spec_comp_L_range} also includes for comparison SN 2006bt,
which displayed Si II 6355 which was higher-velocity and broader than typical SNe Ia, but weaker and lower-velocity than SN 2019yvq.
We would also classify SN 2006bt as a transitional 02es-like
\citep[in agreement with][]{taubenberger_SN_handbook},
and we refer to \citet{foley_06bt} for a thorough study of this unusual object.

02es-like SNe Ia are also characterized by Ti II at peak,
which is seen in lower luminosity SNe Ia like SN 1991bg (see Figure \ref{fig:spec_comp_L_range}).
We note that the spectra of SN 2019yvq and SN 2002bo are quite dissimilar bluewards of $\sim$4500 \AA,
which is precisely at one end of the Ti II ``trough".
Ti II and V II are efficient at suppressing blue flux in SNe Ia, and we refer to Figure 11 of \citet{cartier_15F} to demonstrate their effects on SNe Ia spectra.
In the wavelength regime of the Ti trough, SN 2019yvq is again intermediate between typical-luminosity SNe Ia (SN 2011fe, SN 2002bo) and low-luminosity SNe Ia (SN 2002es, SN 1991bg).
We take SN 2019yvq's suppressed blue flux as tentative evidence for it having Ti, albeit weaker than the more extreme case of SN 1991bg.

Strong [Ca II] and [Fe II] emission is also seen in the nebular spectra of sub-luminous SNe Ia,
such as the 02es-like SN 2010lp \citep{taubenberger_10lp}.
As already discussed in Section \ref{ssec:nebular_spec} and shown in Figure \ref{fig:nebular_spec},
SN 2019yvq displays nebular [Ca II] emission which is intermediate between low-luminosity and normal-luminosity SNe Ia,
again placing it in a transitional region of parameter space.

The weak/nonexistent secondary IR maximum, relatively high decline rate, nebular [Ca II], and Ti II are all pieces of evidence in support of SN 2019yvq being an underluminous event.
When the appropriate extinction is used,
this brings its peak luminosity and color to the border of what could be considered 02es-like SNe Ia, and we classify it as a transitional member of that subclass.

\begin{table}[t!]
\begin{center}
\begin{tabular}{ |c|c|c|c|c| } 
 \hline
 02es-like SN & Host & Earliest & Filter & Early \\
 & type & epoch & & excess? \\
 & & (days) & & \\
 \hline
  SN 2019yvq$^1$ & SAB0 & -15.8 & Swift & Yes \\
  iPTF14atg$^2$  & E-S0 & -15.5 & Swift & Yes \\
  iPTF14dpk$^3$  & Starburst & -16.3 & R    & Maybe \\
  PTF10acdh$^4$ & $\cdots$ & -14.5 & R     & Unknown \\
  PTF10ujn$^4$  & $\cdots$ & -10.7 & R     & Unknown \\
  PTF10bvr$^4$  & E  & ??    & R           & Unknown \\
  SN 2002es$^5$  & S0 & -7.3  & B           & Unknown \\
  SN 1999bh$^6$  & Sb & 0.6   & B           & Unknown \\
  SN 2006bt$^{6,7}$  & S0/a & -2.6     & B           & Unknown \\
  PTF10ops$^{6,8}$  & SAa? & -6.6     & B           & Unknown \\
  SN 2010lp$^6$  & SAb & -7       & B           & Unknown \\
 \hline
\end{tabular}
\caption{
A literature sample of known 02es-like SNe Ia.
iPTF14atg is the only other 02es-like observed in blue filters as early as SN 2019yvq, and it also displays a UV excess.
iPTF14dpk displayed a sharp rise from its last non-detection,
and its first detection is high relative to a power law rise.
PTF10ops is either $\sim$148 kpc offset from the spiral galaxy SDSS J214737.86+055309.3, or in a very faint satellite galaxy of it.
Sources:
1: this work;
2: \citet{Cao15};
3: \citet{Cao16};
4: \citet{white15};
5: \citet{Ganeshalingam12};
6: \citet{taubenberger_SN_handbook};
7: \citet{foley_06bt};
8: \citet{maguire_ptf10ops}.
}
\label{table:02es_litsample}
\end{center}
\end{table}

Table \ref{table:02es_litsample} lists all known 02es-like SNe Ia, including SN 2019yvq.
The three SNe which were detected the earliest all display unusual lightcurve properties.
iPTF14atg \citep{Cao15} has already been discussed as a prime example of an early UV excess.
The early lightcurve of iPTF14dpk \citep{Cao16} differed from iPTF14atg,
as it rose more than 1.8 magnitudes/day between its last non-detection and earliest detection (in $R$, the only observed band at that epoch).
\citet{Cao16} take this as evidence of a dark phase,
a time period after the explosion where the energy generated by radioactive decay has not yet reached the photosphere (i.e. the explosion has occurred but is not yet visible).
The lightcurve also declined between the first and second epochs,
although \citet{Cao16} attribute this to scatter consistent with the errors and not a physical dimming.
The paper concludes that the lightcurve of iPTF14dpk is consistent with the ejecta-companion interaction scenario but seen from an unfavorable viewing angle.

The fact that the three 02es-like SNe Ia which have the earliest observations all display extremely unusual, but consistent, lightcurve properties could be evidence that they all arise from identical progenitor systems,
but the sample of such well-observed events will need to be expanded beyond its current limited numbers to make this statement with statistical confidence.
But even with the small sample size we can say that the companion-ejecta interaction models, which predict a strong UV excess $\sim$10\% of the time due to viewing angle constraints,
are unlikely to be the source of 02es-like SNe Ia if two of the three SNe observed at the right epochs display such an excess with certainty,
and the third displays a potential weak excess.
We discuss these implications more in Section \ref{sec:discussion}.

\section{Model Comparisons}\label{sec:models}

\begin{figure}
\begin{center}
\includegraphics[width=0.5\textwidth]{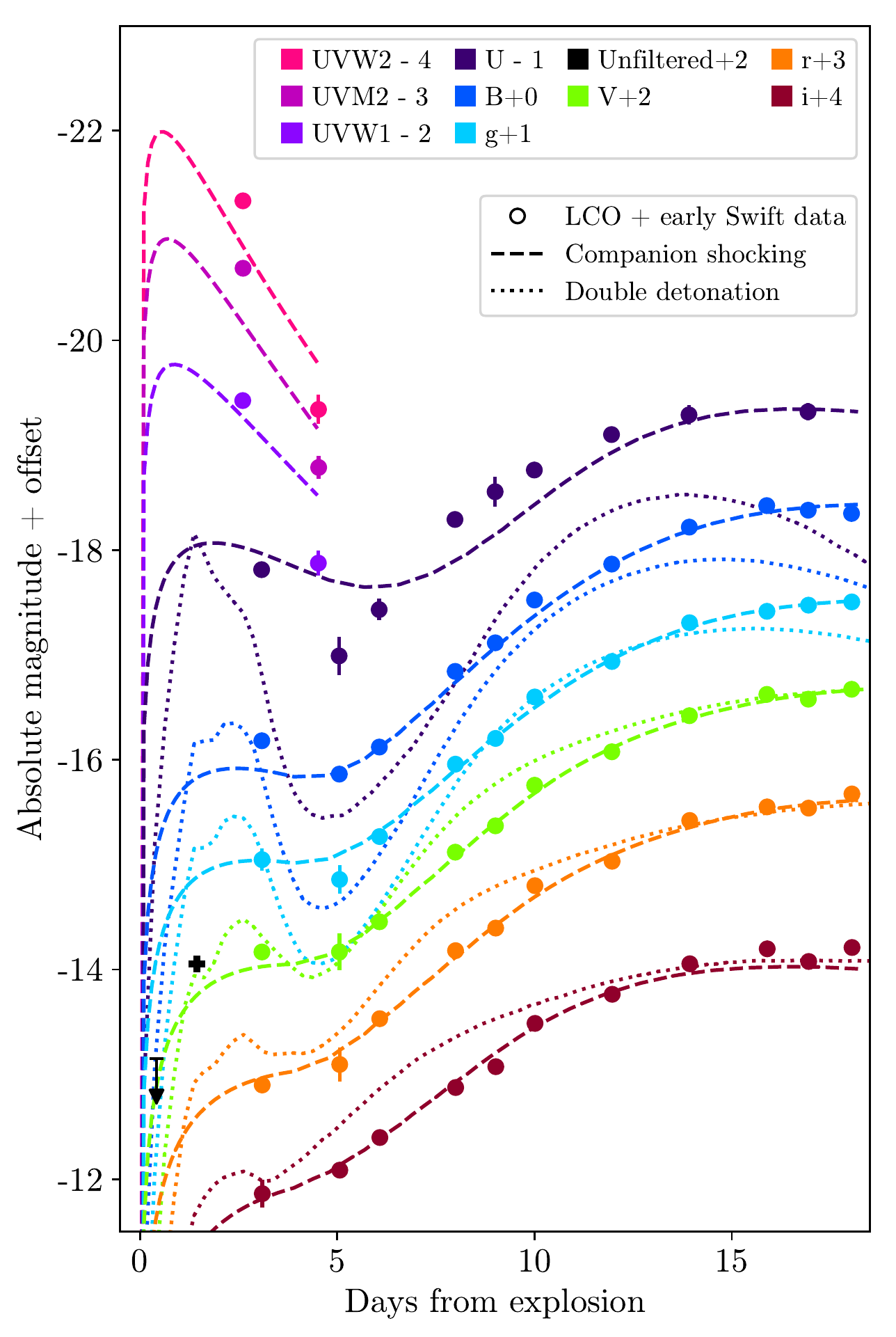}
\caption{Comparisons between the Las Cumbres and early \textit{Swift} data for SN 2019yvq and two different models.
The non-detection and first detection from Itagaki are included in black.
Shown in the dashed line is the best-fit companion shocking model from \citet{kasen10}.
The parameters for this model are in Table \ref{table:kasen_param}
(see Section \ref{ssec:companion_shocking} for more detail).
The SN template used to generate the companion shocking model did not extend into the mid-UV,
so only the blackbody flux component is shown for the \textit{Swift} filters.
The dotted line is the best-fit double detonation model from \citet{polin_subch}:
a 0.95 \msun \ WD progenitor with 0.055 \msun \ of He
(see Section \ref{ssec:double_det} for more detail).
}
\end{center}
\label{fig:model_comp}
\end{figure}

We compare SN 2019yvq to two main classes of models which are capable of producing early blue bumps: companion shocking models from \citet{kasen10} and double detonation sub-Chandrasekhar mass models from \citet{polin_subch}.
Our best-fit models in these two categories are included in Figure \ref{fig:model_comp}.
We also discuss comparisons to models with varying Ni distributions.
No one model reproduces all features of the dataset, so we discuss their benefits and shortcomings.

\subsection{Companion Shocking} \label{ssec:companion_shocking}

As discussed in the introduction,
\citet{kasen10} predicted that an early blue/UV excess could be seen in the lightcurves of SNe Ia when the ejecta collide with a nondegenerate companion and gets shock-heated.
This excess arising from companion shocking would only be visible within a few days of the explosion,
and would only be seen for $\sim$10\% of SNe Ia due to viewing angle effects.

\citet{Hosseinzadeh17} previously used these models to fit the lightcurve of SN 2017cbv.
As described in that paper, they require a total of eight parameters to generate fits:
(1) the explosion epoch $t_0$, 
(2) the companion separation $a$, 
(3) a factor involving the ejecta mass and speed $(x \propto M v^7)$,
(4) the time of maximum $t_{\textrm{max}}$,
(5) the lightcurve stretch $s$,
(6) and (7) factors on the $r$ and $i$ flux of the SiFTO template \citep{sifto} $r_r$ and $r_i$,
and (8) a factor on the $U$ shock flux $r_U$.

We make use of \texttt{lightcurve\_fitting} \citep{griffin_lightcurvefitting} to fit these models,
which uses a Markov Chain Monte Carlo routine based on the \texttt{emcee} package \citep{emcee} to generate fits.
The models consist of two components:
a blackbody flux component and a SiFTO template which can be stretched and scaled.
We extend the blackbody component of the model to include the early 
\textit{UVW2},
\textit{UVM2}, and
\textit{UVW1}
\textit{Swift} data,
since the first two epochs were taken in a regime where the SN flux was dominated by the early excess.

Fits struggled to converge until the following steps were taken:
(1) we put a tight prior on the explosion epoch and enforced adherence to the non-detection from Itagaki Astronomical Observatory, and
(2) we extended the multiplicative factor on the $U$ shock flux to include \textit{Swift} data due to the strength of the excess in those bands as well.
The parameters for our best-fit model are listed in Table \ref{table:kasen_param},
along with the corresponding best-fit model for SN 2017cbv from \citet{Hosseinzadeh17}.

\begin{table}[t!]
\begin{center}
\begin{tabular}{ |c|c|c| } 
 \hline
  & SN 2019yvq & SN 2017cbv \\
 \hline
 $t_0$ (MJD) & 58844.3$\pm$0.1 & 57821.9 \\
 $a$ (\rsun) & $52^{+6}_{-4}$& 56 \\
 $\frac{M}{\textrm{M}_{\textrm{Ch}}}
 \left( \frac{v}{10000 \textrm{ km s}^{-1}} \right)^{7}$ & $0.099\pm0.03$ & $3.84\pm0.19$ \\
 $t_{\textrm{max}}$ (MJD) & $58863.14\pm0.08$ & 57840.2 \\
 $s$    & $0.878\pm0.007$ & 1.04 \\
 $r_r$  & $0.920\pm0.006$ & 0.95 \\
 $r_i$  & $0.736^{+0.006}_{-0.007}$ & 0.85 \\
 $r_U$  & $1.27\pm0.04$ & 0.61 \\
 \hline
\end{tabular}
\caption{
Comparisons between the best-fit parameters of the \citet{kasen10} companion shocking models for SN 2019yvq (this work) and SN 2017cbv \citep{Hosseinzadeh17}.
Parameters: 
time of explosion ($t_0$),
companion separation ($a$),
a parameter involving the ejecta mass and velocity ($\propto M v^{7}$),
time of peak ($t_{\textrm{max}}$),
lightcurve stretch ($s$),
factors on the $r$ and $i$ flux in the SiFTO template ($r_r, r_i$),
and a flux factor on the $U$ though $UVW2$ shock flux ($r_U$).
}
\label{table:kasen_param}
\end{center}
\end{table}

The most significant of these is the $r_U$ factor: \citet{Hosseinzadeh17} find that the $U$ shock flux for models describing SN 2017cbv must be scaled by a factor of 0.61.
There are several possible explanations for this,
including assumptions of spherical symmetry and blackbody SEDs,
or the effects of line blanketing from iron group elements (IGEs) causing the UV/blue flux to be overestimated.

However, we do \textit{not} find that the $U$ (and $UVW1$, $UVM2$, $UVW2$) shock flux needs to be scaled down to match the data.
Instead the best-fit model has a UV flux enhancement of about 27\%.
An increase of this amount is unsurprising: 
the analytic expressions for the blackbody luminosity used in \texttt{lightcurve\_fitting} and derived from \citet{kasen10} replicate the numerical models of companion-ejecta interaction seen at a viewing angle of approximately $30^{\circ}$ (see Figure 2 of that paper).
Explosions with smaller viewing angles result in higher observed luminosities,
up to about 0.25 dex (a factor of 1.8) brighter for a perfectly aligned scenario.
Although our model does not include the viewing angle as a parameter,
better-aligned explosions can generate the required shock flux enhancement.

The other notably discrepant parameter between the two fits is the parameter involving mass and velocity.
It is worth noting that the relevant velocity is not exactly the ejecta velocity,
rather it is the transition velocity between different power laws in the density profile for the modeled ejecta.
Assuming $\textrm{M}_{\textrm{Ch}}$ of ejecta,
the value of this parameter for SN 2017cbv corresponds to a velocity of about 12000 km s$^{-1}$.
Using the same assumption,
the value for SN 2019yvq corresponds to a transition velocity of about 7000 km s$^{-1}$.

The best-fit companion separation (52 \rsun) implies 
a companion radius of $\sim$20 \rsun, assuming Roche lobe overflow \citep{eggleton}.
This stellar radius does not exclude most main sequence stars,
and the separation lies towards the extreme of the expected distribution for main sequence donor stars, based on binary population synthesis models \citep{liu_15}.

\citet{miller2020_19yvq} also use the \citet{kasen10} models to fit their data,
although with a different methodology.
They fit only shock-dominated data (within $\sim$3.5 days of explosion)
and use a slightly different analytical form for the shock flux.
They find a best-fit companion separation of $13\pm1$ \rsun \ 
and an explosion date of $58845.82\pm0.04$ (MJD).
This companion separation is several times smaller than our best-fit value (Table \ref{table:kasen_param}), and the explosion date is more than 1.5 days after ours.
Since their explosion date is in fact almost two hours after the initial detection from Itagaki,
we are unsurprised by the disagreement in companion separations.

As a final remark on the best-fit parameters in Table \ref{table:kasen_param},
we note that SN 2019yvq and SN 2017cbv have similar rise times
(18.7 days and 18.2 days, respectively).
These values are quite typical for SNe Ia -- \citet{Firth15} find an average rise time of $18.98\pm0.54$ days in a sample of 18 well-sampled objects.

\begin{figure}[t]
\includegraphics[width=0.49\textwidth]{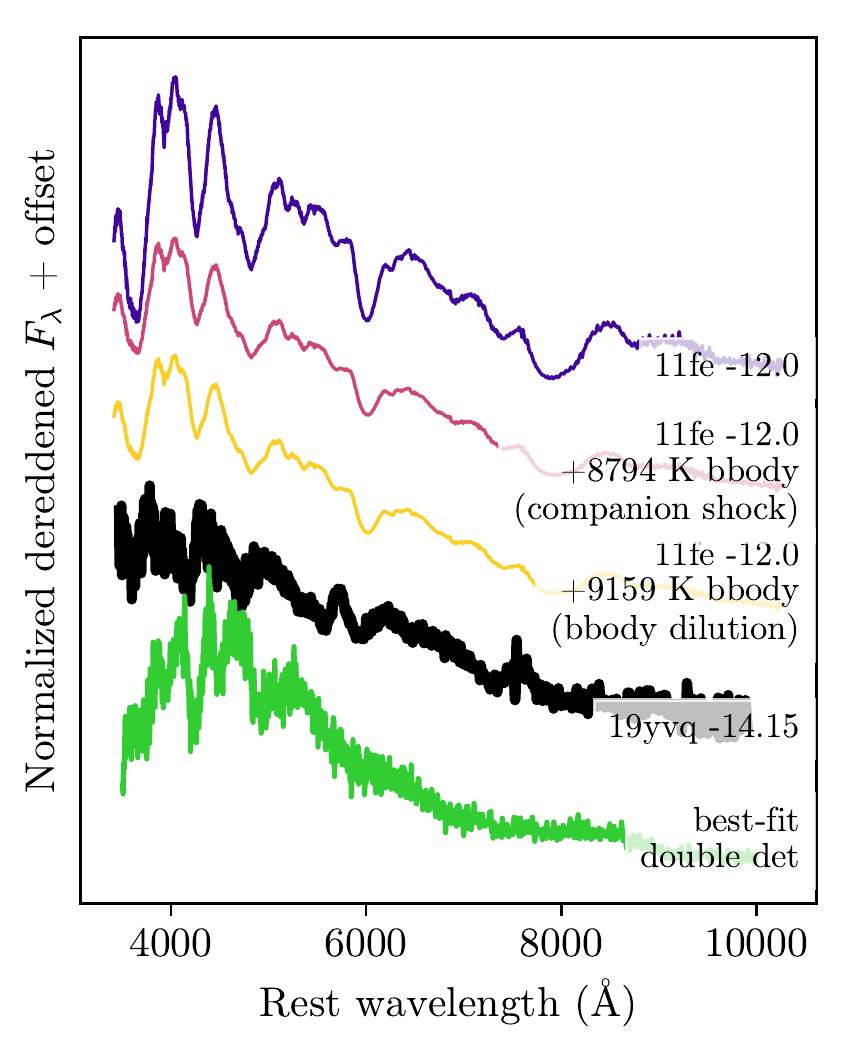}
\caption{ 
Our earliest spectrum of SN 2019yvq (black line) compared to a spectrum of SN 2011fe at a comparable epoch.
Epochs listed with respect to days from \textit{B}-band maximum.
The magenta line represents the SN 2011fe spectrum diluted by a 8794 K blackbody, the temperature predicted at that epoch by our best-fit companion shocking models.
Allowing the temperature of the blackbody to vary and comparing to the the SN 2019yvq with a $\chi^{2}_{\nu}$ test,
we obtain a best-fit temperature of about 350 K higher (yellow line).
The green line represents the spectrum at the same epoch (measured from explosion) from the best-fit double detonation model.}
\label{fig:spec_11fe}
\end{figure}

Although \texttt{lightcurve\_fitting} generates model lightcurves and not spectra,
we reproduce the spectral effects of this model by taking a spectrum of SN 2011fe at a similar epoch to our earliest spectrum and diluting it with a blackbody of the predicted size and temperature.
The effects of this blackbody dilution are shown in Figure \ref{fig:spec_11fe},
where it can be seen that they do a qualitatively good job replicating the early spectrum of SN 2019yvq (in black),
with its blue continuum and weak features.
Further, quantitatively fitting for the best-fit temperature needed to reproduce the strength of spectral features
(keeping the radius the same as predicted by the fits)
results in a temperature only about 350 K higher than predicted by the models.
These two temperatures being consistent with each other provides independent confirmation of the validity of the companion shocking models.

Companion shocking models can produce a wide range of early blue bumps depending on the companion separation, size, and viewing angle \citep[see Figures 2 and 3 of][]{kasen10}.
While the fits for SN 2019yvq are not perfect,
notably underpredicting the strength of the decline to the second epoch of \textit{Swift} data,
they both closely reproduce the wavelength-dependent behavior of the early excess and predict a temperature closely aligned with what is expected by diluting an early spectrum with blackbody flux.

\subsection{Double Detonation} \label{ssec:double_det}

As described in detail in \citet{polin_subch},
the explosion mechanism of these models consists of the ignition of a surface layer of He which then detonates the underlying C/O WD.
We compared observations of SN 2019yvq with double-detonation models which had WD masses between 0.6 and 1.3 \msun \ 
and He shell masses between 0.01 and 0.1 \msun.

We measure the overall best-fit model in our grid by doing a simple reduced $\chi^2$ comparison between each model and the $UBVgri$ photometry.
We fix the explosion epoch to be the same used in the best-fit companion shocking model,
as described in Section \ref{ssec:companion_shocking}.
Normally one would infer an explosion epoch from a power-law fit to the rising data
\citep[e.g.][]{ganeshalingam_powerlawrise, firth_powerlawrise}
however in this case these fits were very poorly constrained.
This was primarily due to a limited number of epochs available for fitting, as there were only four left after ignoring the obviously non-power-law first epoch.

The best-fit model in our grid has a 0.95 \msun \ WD with a 0.055 \msun \ layer of He.
This model is shown as the dotted line in the photometry of Figure \ref{fig:model_comp} and the color evolution of Figure \ref{fig:color},
and the spectrum from this model matching the epoch of our earliest SN 2019yvq spectrum is shown in Figure \ref{fig:spec_11fe}.
Although most of this spectrum is a blue continuum with weak features, in general agreement with the observations,
we find that it predicts much stronger features in the $\sim$4000--5000 \AA \ range and a stronger downturn blueward of $\sim$4000 \AA \ than are observed.

This model does have a strong excess at the correct epochs
(i.e. up to $\sim$4 days after the explosion),
however it dramatically underpredicts most of the \textit{U} data.
The drop after the early excess is also stronger in all bands than is seen in the data,
and the models predict a ``red bump" which is not seen in the data (see Figure \ref{fig:color}).
Additionally, all reasonably well-fitting models in the grid predict a \textit{U} decline that is steeper than observed.
In the case of the best-fit model, it is steeper than the observed decline-rate by more than a factor of two (in magnitudes per day).

There are also several advantages to double detonation models which match the observed data:
a lack of C in the spectra,
a weak secondary IR maximum,
and a blue/UV excess at roughly the right epochs are some points of agreement.

Both \citet{miller2020_19yvq} and \citet{siebert_19yvq} use the models from \citet{polin_subch} to fit different aspects of SN 2019yvq's dataset.
Fitting to the \textit{gri} ZTF photometry in addition to some \textit{Swift} data over approximately the same epochs shown here,
\citet{miller2020_19yvq} find a best-fit model consisting of a 0.92 \msun \ WD with a 0.04 \msun \ He shell.
Their results are similar to what is presented here:
general agreement on some counts (early blue excess),
and diagreement on others (difficulty fitting bluer filters).

\citet{siebert_19yvq} extend the best-fit model of \citet{miller2020_19yvq} into the nebular phase, and show that the best-fit model based on photospheric photometry is a poor match for nebular spectroscopy, 
overpredicting the strength of the [Ca II] and [Fe II] feature by a factor of several.
Instead, to match the nebular spectra they find a best-fit model consisting of a 1.1 \msun \ WD with a 0.05 \msun \ He shell.
This nebular model is in turn a poor match to the photospheric photometry,
overpredicting the bluer bands by more than a magnitude and greatly underpredicting the strength of the early excess in optical bands.

We find it difficult to reconcile this discrepancy, and cannot definitively claim that SN 2019yvq is the result of a double-detonation, despite the several points in favor of these models as listed above.

\subsection{Nickel Distributions} \label{ssec:ni_models}

\subsubsection{Photometry}\label{sssec:ni_phot}

Variations in Ni distributions in the WD progenitor are also known to produce a range of SN Ia behavior \citep[e.g.][]{Piro16, Magee2020}.

Using the same methodology described in Section \ref{ssec:double_det},
we look for best-fit models from the grid of 255 models provided by \citet{Magee2020}.
These models make use of the radiative transfer code \texttt{TURTLS} \citep{Magee2018} and vary the density profiles, Ni masses, kinetic energy, and degree of Ni mixing to produce a range of lightcurves up to $+25$ days from the explosion.

Fitting the  \textit{UBVgri} Las Cumbres lightcurve,
we find the best-fit model is \texttt{EXP\_Ni0.8\_KE0.50\_P4.4}.
This has an exponential density profile,  0.8 \msun \ of Ni, and a kinetic energy of 0.50 foe.
The last element of the model name (\texttt{P4.4}) describes the scaling parameter which determines the Ni distribution,
and indicates the Ni is comparatively mixed through the ejecta.

However, while this model does as well as the other two classes of models we have discussed at fitting the rise time and peak absolute magnitude,
it contains no early excess.
The authors note in \citet{Magee2020}
that although they can fit a majority of SNe in their sample, the remaining objects have an early excess which the models cannot replicate.
Since we consider the early UV excess to be the most unique feature of this SN, the most difficult and interesting aspect to model, and potentially the biggest clue to what the progenitor system is,
we do not include this best-fit model in Figure \ref{fig:model_comp}.

The same authors also released a set of models using a similar methodology capable of reproducing early excesses due to clumps of $^{56}$Ni in the outer ejecta \citep{magee_clumpsandbumps}.
However, since these models were based on SN 2017cbv and SN 2018oh data and both these SNe had typical peak luminosities unlike the underluminous SN 2019yvq, we do not include them as comparisons.
Additionally, these models display early red bumps similar to those seen in the double detonation models,
which are not seen in our data (see Figure \ref{fig:color}).

\subsubsection{Spectroscopy}\label{sssec:ni_spec}

In addition to the above photometric modeling,
we also utilize \textsc{Tardis} \citep{Kerzendorf2014} to examine the spectroscopic effects of varying Ni distributions and photospheric velocities.
A full exploration of these effects are outside the scope of this paper, but we report initial observations here.

We start with a base model,
which consists of an early SN 2011fe spectrum identical to the one used in \citet{heringer2017} at an epoch of $+5.9$ days from the explosion, similar to the epoch of our earliest spectrum.
The \texttt{v\_inner\_boundary} (photospheric velocity) of this model is $12,400$ km/s.
We then alter the Ni distribution and photospheric velocity of this model in an attempt to replicate the SN 2019yvq.

Our perturbations were unsuccessful at reproducing the earliest spectrum, but we note observable effects of altering the Ni distribution.
Adopting a uniform Ni distribution for the outer ejecta with a mass fraction of 0.19 \citep[replicating the most mixed model of][]{Piro16},
we note that the red wings of the Si II 6355 and O I 7774 lines become asymmetrically broader, and that the Ca NIR triplet drastically reduces in strength.
Artificially introducing a mass of Ni in the outermost portions of the ejecta ($>20,000$ km/s) weakens the Mg II complex and other features blueward of $\sim$4500 \AA.
As the density of this outer Ni mass is increased, other dramatic effects, such as the extreme broadening of the O I 7774 features are introduced, which are not seen in the early spectra of SN 2019yvq.

We also experiment with varying the photospheric velocity of the models,
as our earliest spectrum has a Si II 6355 velocity of approximately $21,000$ km s$^{-1}$, which is
significantly higher than the default value of $12,400$ km s$^{-1}$.
\citet{miller2020_19yvq} find velocities of as high as $25,000$ km s$^{-1}$ are necessary to fit their earliest spectrum, 
but since the maximum velocity in the \textsc{Tardis} model is $24,000$ km s$^{-1}$ this is unreachable for us.
We do note that at high photospheric velocities, such as $18,000$ to $20,000$ km s$^{-1}$,
the strengths of most spectroscopic features begin to match the weak values of our earliest spectrum and the spectrum begins to be dominated by a blue continuum.
However, as also pointed out by \citet{miller2020_19yvq},
\textsc{Tardis} has a photospheric boundary which is not wavelength-dependent inside of which is a quasi-blackbody.
Because our \textsc{Tardis} models have a limited velocity range,
increasing the model's photospheric velocity thus increases the percentage of the model's mass which acts as a blackbody and effectively dilutes the spectral features from the tenuous outer layers with a strong blackbody component.
Blackbody dilution is also a signature of the companion shocking models, and is shown in Figure \ref{fig:spec_11fe}.
The blackbody temperature predicted by the companion shocking models is also thousands of Kelvin hotter than the photospheric temperatures \textsc{Tardis} calculates for this velocity range (between 6,000 and 7,000 K).

\citet{miller2020_19yvq} use additional Ni distribution models based on \citet{magee_clumpsandbumps} and find that the predicted spectra have strong line blanketing blueward of $\sim$4400 \AA,
in addition to overpredicting the \textit{i}-band flux.

Since unusual Ni distributions result in spectral features absent in the observed spectra,
and since high photospheric velocities replicate the effects of the companion interaction scenario,
we do not include these spectra in our comparisons.

\section{Progenitor Constraints from Radio Observations} \label{sec:radio}
Radio emission is a sensitive probe of circumstellar medium (CSM) of the progenitor. The CSM is polluted by mass-loss from the progenitor in the pre-SN stage, and interaction of the SN ejecta with this CSM accelerates electrons to relativistic energies and amplifies the ambient magnetic field, producing synchrotron radio emission \citep{Chevalier1982, Chevalier1984, Chevalier1998}. Simple models of radio emission have provided constraints on the CSM environment and progenitor properties for both core-collapse \citep[e.g.][]{Ryder2004, Soderberg2006, Chevalier2006, Weiler2007, Salas2013} and SNe Ia \citep{Panagia2006, Chomiuk2016}. Radio emission is yet to be detected from a SN Ia , but non-detections have provided stringent constraints on progenitor scenarios \citep{Chomiuk2016}, particularly for nearby events like SN\,2011fe \citep{Horesh2012, Chomiuk2012} and SN\,2014J \citep{Torres2014}. 

Radio observation of SN\,2019yvq was obtained with the Karl G. Jansky Very Large Array (VLA) on 2020 Jan 26, 11:39:53, which is within 29.77 days of $t_0$ (derived in Section \ref{ssec:obs_phot}). The observation block was 1-hr long, with 38.23 mins time-on-source for SN\,2019yvq. Observations were taken in X-band (8--12 GHz) in the D-configuration of the VLA (DDT: 19B-346, PI: S. Sarbadhicary).\ The observations were obtained in wide-band continuum mode, yielding 4 GHz of bandwidth sampled by 32 spectral windows, each 128 MHz wide sampled by 1 MHz-wide channels with two polarizations. We used 3C286 as our flux and bandpass calibrator, and J1313+6735 as our phase calibrator. 
Data were calibrated with the VLA CASA calibration pipeline (version 5.6.2-2)
\footnote{\url{https://science.nrao.edu/facilities/vla/data-processing/pipeline}}.
The pipeline consists of a collection of algorithms that automatically loads the raw data into a CASA measurement set (MS) format, flags corrupted data (e.g. due to antenna shadowing, channel edges, radio frequency interference or RFI), applies various corrections (e.g. antenna position, atmospheric opacity) and derives delay, flux-scale, bandpass and phase calibrations which are applied to the data.

We imaged the calibrated visibility dataset with \texttt{tclean} in CASA. We used multi-term, multi-frequency synthesis as our deconvolution algorithm (set with \texttt{deconvolver=`mtmfs'} in \texttt{tclean}), which performs deconvolution on a Taylor-series expansion of the wide-band spectral data in order to minimize frequency-dependent artifacts \citep{Rau2011}. We set \texttt{nterms=2} which uses the first two Taylor terms to create images of intensity (Stokes-I) and spectral index. The SN is offset $\sim 13^{\prime\prime}$ from the bright central radio nucleus of the galaxy, and as a result the emission at the SN site is dominated by sidelobes from the nucleus for the typical resolution $\sim7.2^{\prime\prime}$ expected in X-band images in D-configuration. For this reason, we only imaged the 10-12 GHz bandwidth with \texttt{tclean}, excluded visibility data from baselines shorter than 6 k$\lambda$, and applied Briggs-weighting on the remaining visibility data with the parameter \texttt{robust=0}. This provided just enough angular resolution and source sensitivity at the SN site to determine if any radio emission separate from the nucleus is associated with the SN site.

No radio source was detected at the site of SN 2019yvq in the cleaned, deconvolved 11-GHz image with a synthesized beam of $5.5^{\prime\prime} \times 4.2^{\prime\prime}$. The flux at the exact location of the SN is $-25\mu$Jy. Using the AIPS task \texttt{IMEAN}, we obtain an RMS of $11.7\mu$Jy per beam, which translates to a 3$\sigma$ 11-GHz luminosity limit of 
$7.6 \times 10^{25}$
ergs/s/Hz, assuming a distance of 
42.5 
Mpc. 

\begin{figure}
\begin{center}
\includegraphics[width=0.5\textwidth]{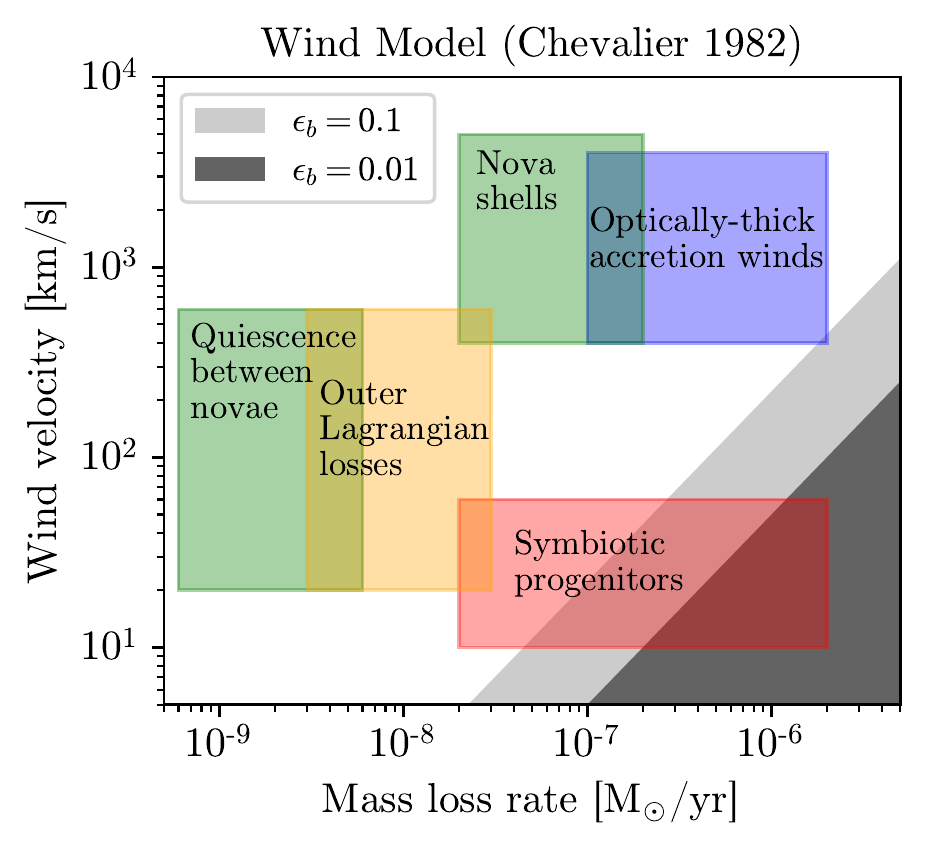}
\caption{
Limits (in gray) for the mass loss rate of the progenitor of SN 2019yvq from its VLA observations,
following the model of \citet{Chevalier1982}, shown for typical range of values of $\epsilon_b$ which parameterizes the fraction of shock energy in the amplified post-shock magnetic field in radio light curve models.
These observations can rule out some symbiotic progenitor systems, but they do not exclude red giant companions or other methods of mass loss.
}
\end{center}
\label{fig:radio_limits}
\end{figure}

The 3$\sigma$ upper limit can shed some light on the CSM around 2019yvq similar to the methodology in \cite{Chomiuk2012} and \cite{Chomiuk2016}. Using the \cite{Chevalier1982} model of a CSM characterized by $\rho = \dot{M}/4 \pi r^2 v_w$ (where $\rho$ is density in gm/cm$^3$, $\dot{M}$ is the mass-loss rate from the progenitor, $r$ is the distance from progenitor and $v_w$ is wind velocity), we obtain an upper limit of 
$(4.5 \text{---} 20) \times 10^{-8}$
M$_{\odot}$/yr on the mass-loss rate from a symbiotic progenitor (involving a red-giant companion, assuming $v_w$=10 km/s). The range of mass-loss rates reflect the uncertainty in the parameter $\epsilon_b$, the fraction of shock energy shared by the amplified magnetic field, with typical values in the range 0.01-0.1 for SNe \citep{Chomiuk2012}.
These limits are shown in Figure \ref{fig:radio_limits}.
\cite{Chomiuk2016} measured the mean mass-loss rate in symbiotic progenitors in the Milky Way to be $\mathrm{log}_{10} (\dot{M}) = -6.41 \pm 1.03$ M$_{\odot}$/yr (asssuming $v_w=100$ km/s), so our measurement does not exclude the possibility of a red-giant companion. Scenarios involving accretion from a main-sequence companion accompanied by steady nuclear burning are also not excluded by our limit \citep{Chomiuk2012}.

\section{Discussion}\label{sec:discussion}

SN 2019yvq is an unusual event in many respects.
It has:
a strong early UV flash;
red colors besides the early flash;
relatively faint peak luminosity, a moderately high decline rate, and a weak secondary IR maximum;
broad, high-velocity Si II 6355 paired with both weak Si II 5972 and Ti II at peak;
and nebular [Ca II] and [Fe II].
These paint a conflicting picture,
with some aspects pointing to a low-energy explosion
(low luminosity, weak secondary IR maximum, nebular [Ca II], peak Ti II)
and others pointing to a high-energy event (Si II velocity and line ratio).
Due to several characteristics it shares, or almost shares, with low-luminosity 02es-like SNe Ia, we classify it as a transitional member of that subclass (see Table \ref{table:02es} and the rest of Section \ref{sec:02es_comp}).

This object being a transitional 02es-like has two major implications.

The first is the confirmation that transitional 02es-like SNe Ia can exist.
This has precedent in the object SN 2006bt \citep{foley_06bt, ganeshalingam2010},
which can be considered a transitional member of this class \citep{taubenberger_SN_handbook}
despite its high velocities (12,500 km s$^{-1}$ at 3 days before maximum) and relatively bright luminosity ($M_{B,\textrm{peak}}\sim-19$, with uncertain reddening correction).
This object is included in both Figure \ref{fig:spec_demographics} (orange star)
and Figure \ref{fig:spec_comp_L_range} for comparison.
However, SN 2019yvq is by no means a clone of SN 2006bt as it lies in extremely sparsely populated regions of parameter space in several respects \citep[see Figure \ref{fig:spec_demographics}, also Figure 2 of][]{tucker_19yvq}.
On the Phillips relation SN 2019yvq has similar parameters to SN 2012Z,
but on the Branch diagram SN 2019yvq is most similar to SN 2002bo.
SNe 2002bo and 2012Z are substantially different SNe.
A transitional 02es-like SN that not only shares characteristics with both these SNe
but is also distinct from another transitional member of its subclass
supports evidence that 
there is a continuum of events between normal SNe Ia and 02es-likes.
Assuming a continuum of events instead of discrete subclasses,
this also suggests that
02es-like SNe 
do not arise from progenitor systems which are distinct from the systems of normal SNe Ia.

The second major implication comes from the fact that the three 02es-like SNe Ia with very early data (SN 2019yvq, iPTF14atg, and iPTF14dpk) all display unusual early-time lightcurves (see Section \ref{sec:02es_comp} and Table \ref{table:02es_litsample}).
Of these, the two with \textit{Swift} data at these early epochs display the two strongest early UV flashes in SNe Ia.
iPTF14dpk unfortunately only has \textit{R}-band photometry,
and while at first glance its first data point appears indicative of an early excess,
\citet{Cao16} say that this would require an extreme explosion energy and would lead to higher velocities than are observed. 
The lack of multi-band photometry makes us hesitant to accept that conclusion incontrovertibly.
According to \citet{kasen10}, if such early excesses are due to companion--ejecta shock interaction they should only be seen in $\sim$10\% of events with such early data.
Instead, for 02es-like SNe Ia, they are seen in two (or three) of the three early events.
This is unlikely -- even with the current small sample size, the odds of so many early excesses are somewhere between 1 in 100 and 1 in 1000.
And as discussed in Section \ref{ssec:double_det}, the discrepancies between photospheric and nebular best-fit models make us hesitant to claim that SN 2019yvq is a double detonation event either,
even though those models can produce early UV excesses.
We are left considering progenitor scenarios which could produce an early excess which is both fit relatively successfully by shock interaction models but is not viewing angle-dependent.

In addition to models which have already been discussed (double detonations and varied Ni distributions,
see Sections \ref{ssec:double_det} and \ref{sssec:ni_phot}),
there are a few possibilities for progenitor systems configured in such a way to produce more isotropic shocks.
One option lies in the accretion disks which form as the (primary) WD accretes matter.
\citet{leavnon_disks} model the exquisitely sampled early bump seen in the K2 data of SN 2018oh as the interaction of the SN ejecta with what they refer to as ``Disk-Originated Matter,"
since accretion disks could also give rise to bipolar jets.
The addition of an accretion disk and jets would more easily account for the ubiquity of early excesses since these components can be seen more isotropically.
\citet{Piro16}, in addition to modeling the degree of Ni mixing in WD progenitors, also investigate the effects of a more general distribution of CSM.
These models can produce early excesses which occur on a range of timescales and intensities, depending on the total amount of external matter in the CSM and its density scaling.
In particular they can produce early bumps which only last $\sim$2 days,
which could explain the (potential) extremely brief excess seen in iPTF14dpk.
These CSM models also get redder immediately after the explosion instead of bluer like the Ni mixing models.
This early reddening more accurately reflects the color evolution of SN 2019yvq.

\citet{Cao16} model the 02es-like SNe Ia iPTF14atg and iPTF14dpk as interacting with non-degenerate companions, but seen from different viewing angles.
The addition of SN 2019yvq as another member of the rare 02es-like subclass, with a commensurate early UV excess,
leads us to doubt that all three of these excesses arise from ejecta--companion shock interaction.
Something about their progenitor systems must be more isotropic than is assumed in \citet{kasen10} to explain the ubiquity of these early excesses in 02es-like SNe Ia.


\section{Conclusions \& Summary}\label{sec:conclude}

We have discussed the discovery and follow-up observations of SN 2019yvq,
a nearby SN Ia with a rare and unusually strong excess in its early lightcurve, in addition to several other uncommon features.
This early excess is most pronounced in the UV,
where the object is brighter during the excess than during the epochs of its optical peak.

This object is one of a very limited number of SNe Ia with early UV/blue excess,
and it demonstrates an even stronger excess than other objects in the sample.
SN 2019yvq deviates significantly from SNe Ia that are blue at early times but otherwise normal.
Instead it shares some, but not all, features of the 02es-like SN Ia subclass, including a low peak luminosity, red color, moderately high decline rate, Ti II at peak, and nebular [Ca II] and [Fe II].
We classify SN 2019yvq as a transitional member of the 02es-like subclass.

Although models which simulate WD double detonation and ejecta--companion shock interaction can create lightcurves with excess flux at early times,
we find that no one model can accurately reproduce all unusual aspects of this object's dataset.
This is in broad agreement with the conclusions drawn in \citet{miller2020_19yvq} and \citet{tucker_19yvq},
which include several pieces of data not present here
(including \textit{i}-band ZTF data, post-maximum TESS data, and a Keck NIRES spectrum)
and, like us, are unable to satisfactorily explain every aspect of the SN 2019yvq dataset.
As in \citet{siebert_19yvq} we also find strong [Ca II] and [Fe II] emissions in the nebular spectra of SN 2019yvq in addition to strong limits on the amount of swept-up H and He,
but we do not take this as exclusive evidence of a double detonation explosion.

Two other 02es-like SNe Ia also display unusual early lightcurves (iPTF14atg and iPTF14dpk).
The deviations from a power-law rise in all 02es-like SNe Ia with sufficiently early data makes us further doubt that the early UV excess seen in SN 2019yvq arises from ejecta--companion shock interaction,
as viewing angle effects dictate that such excesses should only be seen in $\sim$10\% of events with early data,
not $\sim$100\%.
02es-like SNe Ia must originate in progenitor systems capable of displaying early excesses nearly isotropically.
The addition of CSM or accretion disks and jets could account for this needed isotropy.

This SN demonstrates the importance of prompt discovery, reporting, and follow-up of young SNe.
In this case, the one day non-detection enabled rapid follow-up with multiple facilities around the world and in space.
The synthesis of such high-cadence multiwavelength datasets is a powerful tool for understanding the origins of SNe Ia,
or for providing even more observational peculiarities which accurate models must account for.

\acknowledgments

We are grateful to A. Polin for providing the lightcurve and spectra models in \citet{polin_subch},
and to G. Hosseinzadeh for assistance in our use of \texttt{lightcurve\_fitting}.
We also thank E. Heringer for providing the \textsc{Tardis} models from \citet{heringer2017},
and R. Cartier for providing the \texttt{syn++} models from \citet{cartier_15F}.

J.B., D.A.H., D.H., C.M., and C.P. are supported by NSF grants AST-1313484 and AST-1911225, as well as by NASA grant 80NSSC19kf1639.

S.K.S. and L.C. are supported by NSF grant AST-1907790.

Time domain research by D.J.S. is supported by NSF grants AST-1821987, 1813466, \& 1908972, and by the Heising-Simons Foundation under grant \#2020-1864. 

P.J.B. is supported by NASA grants 80NSSC20K0456 and 80NSSC19K0316.

This research makes use of observations from the Las Cumbres Observatory network,
in addition to the MARS ZTF alert broker developed by Las Cumbres Observatory software engineers.

This research has made use of the NASA/IPAC Extragalactic Database (NED) which is operated by the Jet Propulsion Laboratory, California Institute of Technology, under contract with NASA.

This research made use of \textsc{Tardis}, a community-developed software
package for spectral synthesis in supernovae
\citep{Kerzendorf2014, tardis_software}.
The development of \textsc{Tardis} received support from the
Google Summer of Code initiative
and from ESA's Summer of Code in Space program. \textsc{Tardis} makes
extensive use of Astropy and PyNE.

%

\vspace{5mm}
\facilities{Las Cumbres Observatory (Sinistro), FTN (FLOYDS), Bok (B\&C Spectrograph), MMT (Blue Channel spectrograph), IRTF (SpeX), Swift (UVOT), VLA}


\software{
\texttt{astropy } \citep{2013A&A...558A..33A,astropy},
\texttt{SNooPy} \citep{snpy},
\textsc{Tardis} \citep{tardis_software},
sncosmo \citep{sncosmo},
SALT2 \citep{salt2},
MLCS2k2 \citep{mlcs2k2},
\texttt{lightcurve\_fitting} \citep{griffin_lightcurvefitting},
\texttt{emceee} \citep{emcee}
          }

\bibliographystyle{aasjournal}
\bibliography{biblio,mlgbib,radio}







\end{document}